\documentclass[usenatbib]{mnras}
\usepackage{lscape}
\usepackage{amsmath,amssymb}
\usepackage{mathrsfs}	
\usepackage{multicol}
\usepackage[dvipdfmx]{graphicx}
\usepackage{textcomp}
\usepackage{txfonts}
\usepackage[T1]{fontenc}
\usepackage{aecompl}

\catcode`\@=11
\def\gta{\ifmmode{\,\mathrel{\mathpalette\@versim>\,}}
    \else{$\,\mathrel{\mathpalette\@versim>}\,$}\fi}
\def\lta{\ifmmode{\,\mathrel{\mathpalette\@versim<\,}}
    \else{$\,\mathrel{\mathpalette\@versim<}\,$}\fi}
\def\@versim#1#2{\lower 2.9truept \vbox{\baselineskip 0pt \lineskip
    0.5truept \ialign{$\m@th#1\hfil##\hfil$\crcr#2\crcr\sim\crcr}}}
\def\fracj#1#2{{\textstyle\frac{#1}{#2}}}
\catcode`\@=12  

\makeatletter
\def\blfootnote{\xdef\@thefnmark{}\@footnotetext}
\makeatother

\newcommand{\vx}{{\bf x}}
\newcommand{\vv}{{\bf v}}
\newcommand{\vq}{{\bf q}}
\newcommand{\vp}{{\bf p}}
\newcommand{\vJ}{{\bf J}}
\newcommand{\Ji}{J_i}
\newcommand{\Jr}{J_r}
\newcommand{\Jphi}{J_{\phi}}
\newcommand{\Jz}{J_z}

\newcommand{\fJ}{f({\bf J})}
\newcommand{\fJi}{f_i({\bf J})}

\newcommand{\Mi}{M_i}
\newcommand{\xperp}{{\bf x}_{\perp}}
\newcommand{\xparal}{x_{||}}
\newcommand{\vperp}{{\bf v}_{\perp}}
\newcommand{\vparal}{v_{||}}

\newcommand{\Mst}{M_{\star}}
\newcommand{\fst}{f_0}
\newcommand{\fJst}{f_{\star}({\bf J})}
\newcommand{\kJ}{k({\bf J})}
\newcommand{\Lz}{L_z}

\newcommand{\xib}{\boldsymbol \xi}
\newcommand{\etaphi}{\eta_{\phi}}
\newcommand{\etaz}{\eta_z}

\newcommand{\Jst}{J_{0,\star}}
\newcommand{\rhos}{\rho_{0, \star}}
\newcommand{\sigmas}{\Sigma_{0, \star}}
\newcommand{\rst}{r_{0,\star}}
\newcommand{\vst}{v_{0,\star}}

\newcommand{\rc}{r_{\rm c, \star}}
\newcommand{\rh}{r_{\rm h}}
\newcommand{\Reff}{R_{\rm e}}
\newcommand{\rhost}{\rho_{\star}}

\newcommand{\fiso}{f_{\rm iso}({\bf J})}

\newcommand{\fJdm}{f_{\rm dm}({\bf J})}
\newcommand{\gdm}{g_{\rm dm}}
\newcommand{\gNFW}{g_{\rm NFW}({\bf J})}
\newcommand{\hJ}{h({\bf J})}
\newcommand{\gcJ}{g_{\rm c}({\bf J})}
\newcommand{\hphi}{h_{\phi}}
\newcommand{\hz}{h_z}
\newcommand{\TJ}{T({\bf J})}
\newcommand{\Jtdm}{J_{\rm t, dm}}
\newcommand{\Jcdm}{J_{\rm c, dm}}

\newcommand{\Mdm}{M_{\rm dm}}
\newcommand{\Jdm}{J_{0,\rm dm}}
\newcommand{\rdm}{r_{0,\rm dm}}
\newcommand{\vdm}{v_{0,\rm dm}}

\newcommand{\Jctil}{\widetilde{J}_{\rm c, dm}}
\newcommand{\Jttil}{\widetilde{J}_{\rm t, dm}}
\newcommand{\Mdmtil}{\widetilde{M}_{\rm dm}}
\newcommand{\Jdmtil}{\widetilde{J}_{0,\rm dm}}
\newcommand{\rhodm}{\rho_{\rm dm}}
\newcommand{\rcdm}{r_{\rm c, dm}}
\newcommand{\rsdm}{r_{\rm s, dm}}
\newcommand{\rtdm}{r_{\rm t, dm}}
\newcommand{\gammadm}{\gamma_{\rm dm}}
\newcommand{\vcdm}{v_{c, \rm dm}}
\newcommand{\MBH}{M_{\rm BH}}
\newcommand{\muBH}{\mu_{\rm BH}}
\newcommand{\Rinf}{R_{\rm infl}}

\newcommand{\PhiBH}{\Phi_{\rm BH}}

\newcommand{\MTL}{\varUpsilon_{\star}}
\newcommand{\chiBHB}{\chi^2_{\rm BHB}}
\newcommand{\chiRHB}{\chi^2_{\rm RHB}}

\newcommand{\alphaBHB}{\alpha^{\rm BHB}}
\newcommand{\alphaRHB}{\alpha^{\rm RHB}}

\newcommand{\etaBHB}{\eta^{\rm BHB}}
\newcommand{\etaRHB}{\eta^{\rm RHB}}

\newcommand{\MstBHB}{M_{\star}^{\rm BHB}}
\newcommand{\MstRHB}{M_{\star}^{\rm RHB}}
\newcommand{\MstRGB}{M_{\star}^{\rm RGB}}
\newcommand{\JstBHB}{J_{\star, 0}^{\rm BHB}}
\newcommand{\JstRHB}{J_{\star, 0}^{\rm RHB}}

\newcommand{\LV}{L_{\rm V}}
\newcommand{\dd}{\text{d}}
\newcommand{\DD}{\partial}

\newcommand{\Msun}{\,M_{\odot}}
\newcommand {\kms} {\,{\rm km\,s}^{-1}}

\begin{document}

\date{Resubmitted, 17 April 2019}
\title[]{Action-based models for dwarf spheroidal galaxies and globular clusters}
{}
\author[]{Raffaele Pascale$^{1,2}$\thanks{E-mail: raffaele.pascale2@unibo.it}, James Binney$^{3}$, Carlo Nipoti$^{1}$ and Lorenzo Posti$^{4}$
\\ \\
$^{1}$Dipartimento di Fisica e Astronomia, Universit\`a di Bologna, via Piero Gobetti 93/2, I-40129 Bologna, Italy \\
$^{2}$INAF Osservatorio di Astrofisica e Scienza dello Spazio, Via Piero Gobetti 93/3, I-40129 Bologna, Italy \\
$^{3}$Rudolf Peierls Centre for Theoretical Physics, Clarendon Laboratory, Oxford OX1
3PU, United Kingdom \\
$^{4}$Kapteyn Astronomical Institute, University of Groningen, P.O. Box 800, 9700 AV Groningen, The Netherlands}

\maketitle

\begin{abstract}
A new family of self-consistent DF-based models of stellar systems is explored.
The stellar component of the models is described by a distribution function (DF)
depending on the action integrals, previously used to model the Fornax dwarf 
spheroidal galaxy (dSph). 
The stellar component may cohabit with either a dark halo, also described by a DF, or
with a massive central black hole. In all cases we solve for the model's
self-consistent potential. Focussing on spherically symmetric models, 
we show how the stellar observables vary with the anisotropy prescribed 
by the DF, with the dominance and nature of the dark halo, and
with the mass of the black hole. We show that precise fits to the observed
surface brightness profiles of four globular clusters can be obtained for a
wide range of prescribed velocity anisotropies. We also obtain precise fits
to the observed projected densities of four dSphs. Finally, we present a 
three-component model of the Scupltor dSph with distinct DFs for the red
and blue horizontal branch stars and the dark matter halo.
\end{abstract}
\begin{keywords}
dark matter - galaxies: dwarf - galaxies: kinematics and dynamics - galaxies: structure - globular clusters: general
\end{keywords}

\section{Introduction}
\label{sec:int}


Diagnosing the dynamics of collisionless systems is central to contemporary
astrophysics. The systems of interest range from clusters of galaxies,
through giant elliptical galaxies and disc galaxies like the Milky Way, to
Magellanic and spheroidal dwarf galaxies and star clusters. All these systems
are dominated by the mass contributed by some mixture of dark-matter
particles and galaxies and stars, and have relaxation times that greatly exceed
their crossing times.  In every case comparison with observations requires
one to recognize that these particles fall into distinct classes: a cluster
of galaxies contains dark-matter particles, and galaxies of several
morphological types; a giant elliptical galaxy contains dark-matter particles
and populations of stars with distinct chemistry; the Milky way and dwarf
spheroidal galaxies (hereafter dSphs) contain dark-matter particles and
populations of stars of distinct chemistry and age, and a globular cluster
(hereafter GC) contains stars with radically different masses and subtly
different chemistry. 

Until recently, dynamical models of stellar systems have been simplified to
the extent of containing only one stellar population and have represented
dark matter by a simple density distribution without regard to its internal
dynamics. With the advent of high-resolution kinematics for 
billions of stars (\citealt{Gaia2018}) and spectra for millions of stars 
(\citealt{Lamost2012}, \citealt{Gaia2018}) it has become essential to develop
multi-component models of stellar systems. In such a model each observationally 
distinct population is represented by a distribution function (hereafter DF) 
$f(\vx,\vv)$ that gives the probability density for finding an object of the
relevant population at the phase-space point $(\vx,\vv)$. Given these DFs,
one can solve for the gravitational potential $\Phi(\vx)$ that these populations 
jointly generate. That done, the model predicts both the spatial distribution 
of each population and the population's velocity distribution at every
point.  

The parameters characterising each component DF can be fitted to data in a
variety of ways. If individual particles are observed, as in a dwarf
spheroidal galaxy, the likelihood of the data given each model and the
observational uncertainties can be computed and used to find the range of
parameters that is consistent with the data \citep[e.g.][]{Pascale2018}.  If
individual particles are not observationally resolved, as in distant
galaxies, the model's parameters can be constrained by comparing observed
surface densities and velocity moments with the model's precisely equivalent
predictions. If the number of resolved particles is large, the cost of
computing individual likelihoods may be unfeasible, forcing one to bin the
data and constrain parameters as in the case of unresolved particles
\citep[e.g.][]{ColeBinney2017}.  Whatever the scale and completeness of the data, a
rigorous and tractable method of parameter constraint is available. 


Models based on a DF have been considered since the beginning of stellar
dynamics (\citealt{Eddington1915}, \citealt{Michie1963}, \citealt{King1966}). 
These models almost invariably take advantage of the Jeans theorem to
posit that the DF depends on $(\vx,\vv)$ only through constants of stellar
motion. The energy $E$ is the most available such constant and until 
recently it invariably featured as an argument of the DF. The key to producing
multi-component and non-spherical models, however, proves to be to exclude
$E$ from the DF in favor of the action integrals $\Ji$ (\citealt{Binney2010}, 
\citealt{Binney2014}). A complete set of action integrals $\Jr$, $\Jz$ and 
$\Jphi$ is guaranteed in any spherical potential, and numerical experiments 
\citep{BinneySpergel1982, Ratcliff1984} with galaxy-like potentials indicate 
that in realistic potentials the vast majority of orbits are quasi-periodic,
which guarantees the existence of action integrals (\citealt{Arnold1978}).
Moreover, by torus mapping (\citealt{BinneyMcMillan2016}) one can closely
approximate any given axisymmetric Hamiltonian with one in which all orbits
are quasi-periodic. Hence it is intellectually sound to require that the DF
depends only on actions.

None the less, it is practicable to take the DF to depend on actions only if
their values can be computed from $(\vx,\vv)$. When \cite{Binney2010} first 
started experimenting with DFs $\fJ$, he used the adiabatic approximation to
compute actions. This approximation works well only for thin-disc stars and
is inapplicable to halo stars or dark-matter particles. Fortunately a
technique for the evaluation of actions soon appeared that provides good
accuracy for all stars and dark-matter particles. This is the `St\"ackel
Fudge' (\citealt{Binney2012a}), which involves using for an arbitrary 
potential formulae that are strictly valid only for St\"ackel's 
separable potentials (\citealt{Stackel}). Recently \cite{Vasiliev2018} 
has released a numerical implementation of the St\"ackel Fudge that is 
highly optimised for speed and is complemented by efficient code for 
solving Poisson's equation for the potential generated by an arbitrary 
axisymmetric mass distribution. \cite{SandersBinney2016} extended the 
St\"ackel Fudge to non-axisymmetric potentials that have no figure rotation.

Early applications of action-based DFs were restricted to modelling the
kinematics of solar-neighborhood stars in given Galactic potentials 
(\citealt{Binney2010}, \citealt{BinneyMcMillan2011}, \citealt{Binney2012b}, 
\citealt{BovyRix2013}). The arrival of the St\"ackel Fudge opened the way 
for global modelling, including imposition of the self-consistency 
condition. \cite{Binney2014} generalised the
isochrone model (\citealt{Henon1960}) to flattened systems, and \cite{PifflPenoyreBinney2015} 
presented a model disc galaxy in which populations of stars spanning a
range of ages self-consistently generate the
potential jointly with a realistic population of dark-matter particles.
Using models of the Fornax dSph in which the potential is
self-consistently generated by stars and dark matter, \cite{Pascale2018}
ruled out the possibility that the phase-space distribution of the
dark matter at the centre of this dark-matter dominated system has the cuspy
structure that is predicted by cosmological simulations that contain only
dark matter.

Central to the art of modelling stellar systems with action-based DFs is a
library of analytic functions $\fJ$ that can be employed for the DFs of
individual components. \cite{Binney2010} introduced a form of the
\textquoteleft quasi-isothermal\textquoteright \, DF, which, refined by
\cite{BinneyMcMillan2011}, has been extensively used to model our Galaxy's
discs. \cite{Posti2015} and \cite{Williams2015a} introduced a family of DFs
$\fJ$ that yield self-consistent models that have two-power-law density
profiles which, inter alia, can closely match the models of \cite{Jaffe1983},
\cite{Hernquist1990} and \citet[][NFW]{NavarroFrenkWhite1996}.
\cite{ColeBinney2017} introduced a modification of the \cite{Posti2015} DFs
that flattens the model's central cusp into a core by making the central
phase-space density finite. 

To model a dSph, \cite{Pascale2018} had to introduce a DF
$\fJ$ that produces systems with exponential rather than power-law outer
density profiles. The purpose of this paper is to explore in a general way
models in which the stellar component is represented by this DF. 
In Section~\ref{sec:fjmodel} we establish our notation.
Section~\ref{subsec:oneC} we explore the dependence of the observable
properties of single-component models on the DF's parameters. In
Section~\ref{subsec:twoC} we embed these models in a dark halo and explore
the dependence of the observables on the degree of dark-matter domination. In
Section~\ref{subsec:BH} we add central massive black holes to the models. In
Section ~\ref{subsec:GC} we show that the density profiles of several well
observed globular clusters can be accurately fitted by the models. For each
cluster we display four models that differ markedly in their kinematics. In
Section~\ref{subsec:dSphs} we use the new DF to fit observations of the dSphs
Carina, Leo I, Sculptor, Sextans and Ursa Minor. In the case of Sculptor our
model assigns distinct phase-space distributions to two populations of
observationally distinguishable stars and a separate component to the dark matter
halo. Section~\ref{sec:conc} concludes.

\section{f({\bf J}) models with multiple components}
\label{sec:fjmodel}

Throughout this paper DFs are normalised to have unit integral over 
phase space:

\begin{equation}\label{for:unity}
\int\dd^3\vq\,\dd^3\vp\,f=1,
\end{equation}
where $(\vq,\vp)$ is any system of canonical coordinates. Let $\fJi$ be such
a DF for  the $i$-th component of a composite stellar system. Sometimes we require
a system's luminosity density, at other times we require its mass
density. Any such phase-space density can be obtained by multiplying $f$ by an
appropriate dimensional factor $Q$; for example, to obtain the dark-matter 
mass density we multiply the DF of dark matter by the total dark-matter mass, and to
obtain the $g$-band luminosity density of a stellar component we multiply
$f_i$ by the component's total $g$-band luminosity.

The real-space mass densities are
\begin{equation}\label{for:rho}
 \rho_i({\bf x}) =\Mi \int\dd^3\vv\, \fJi.
\end{equation}
 The line-of-sight velocity distributions (hereafter LOSVDs) are 
\begin{equation}\label{for:LOSVD}
 \mathcal{L}_i(\xperp, \vparal) = \Mi\frac{\int\text{d}^2\vperp 
 \text{d}\xparal\, \fJi}{\int\text{d}\xparal\,\rho_i(\vx)},
\end{equation}
where $||$ and $\perp$ denote components parallel and orthogonal to
the line-of-sight.

Evaluation of equations (\ref{for:rho}) and (\ref{for:LOSVD}) requires the
mapping between (${\bf x}$, ${\bf v}$) and (${\boldsymbol \theta}$, ${\bf
J}$), which depends on the model's gravitational potential $\Phi$, which is related to
$\rho_i$ via the Poisson equation $\nabla^2\Phi = 4\pi
G\sum_{i=0}^{N}\rho_i$, with $G$ the gravitational constant. We rely on the
St\"ackel-Fudge as implemented in the software library `Action-based galaxy
modelling architecture' (AGAMA\footnote{https://github.com/GalacticDynamics-Oxford/Agama}) that is
described in \cite{Vasiliev2018}, where one can find an extensive analysis 
of the extent to which action values vary along numerically integrated orbits.
The variation exceeds  $\sim2$ per cent only on orbits that have been trapped by a
resonance. We use AGAMA  additionally to solve for self-consistently generated
potentials and to compute moments of DFs. AGAMA provides an optimized interative 
procedure to construct a self-consistent solution, which takes at most less than 
four minutes using an eight core machine (for details, see \citealt{Vasiliev2018}).

\section{Distribution functions for dwarf spheroidals and globular clusters}
\label{sec:dfintro}

\begin{figure*}
 \begin{minipage}{0.33\linewidth}
 \vspace{1.cm}
  \includegraphics[width=1.\hsize]{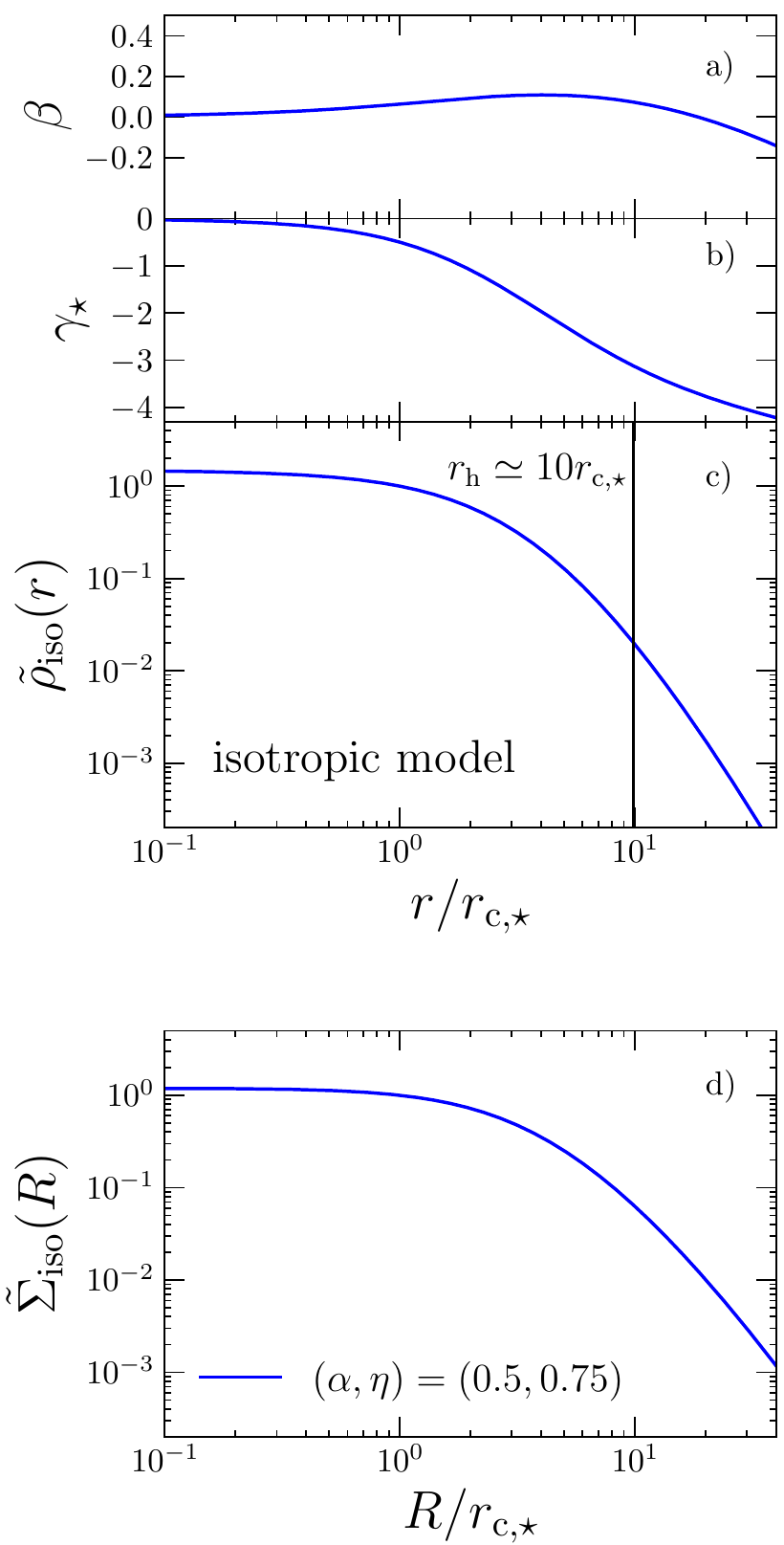}
  \caption{Reference, one-component, isotropic model ($\alpha=0.5$, $\eta=0.75$). From
  top to bottom, anisotropy parameter, slope of the logarithmic density, density and
  projected density are plotted against radius. In the bottom two panels
  $\widetilde{\rho}_{\rm iso}(r)\equiv\rho_{\rm iso}(r)/\rho_{\rm iso}(\rc)$ and 
  $\widetilde{\Sigma}_{\rm iso}(R)\equiv\Sigma_{\rm iso}(R)/\Sigma_{\rm iso}(\rc)$, 
  where $\rho_{\rm iso}$ and $\Sigma_{\rm iso}$ are, respectively, the density 
  and projected density and $\rc$ is the core radius, such that $\gamma_{\star}(\rc)=
  -\frac{1}{2}$.}\label{fig:isotropic} 
 \end{minipage}
 \hfill
 \begin{minipage}{0.65\linewidth}
  \includegraphics[width=1\hsize]{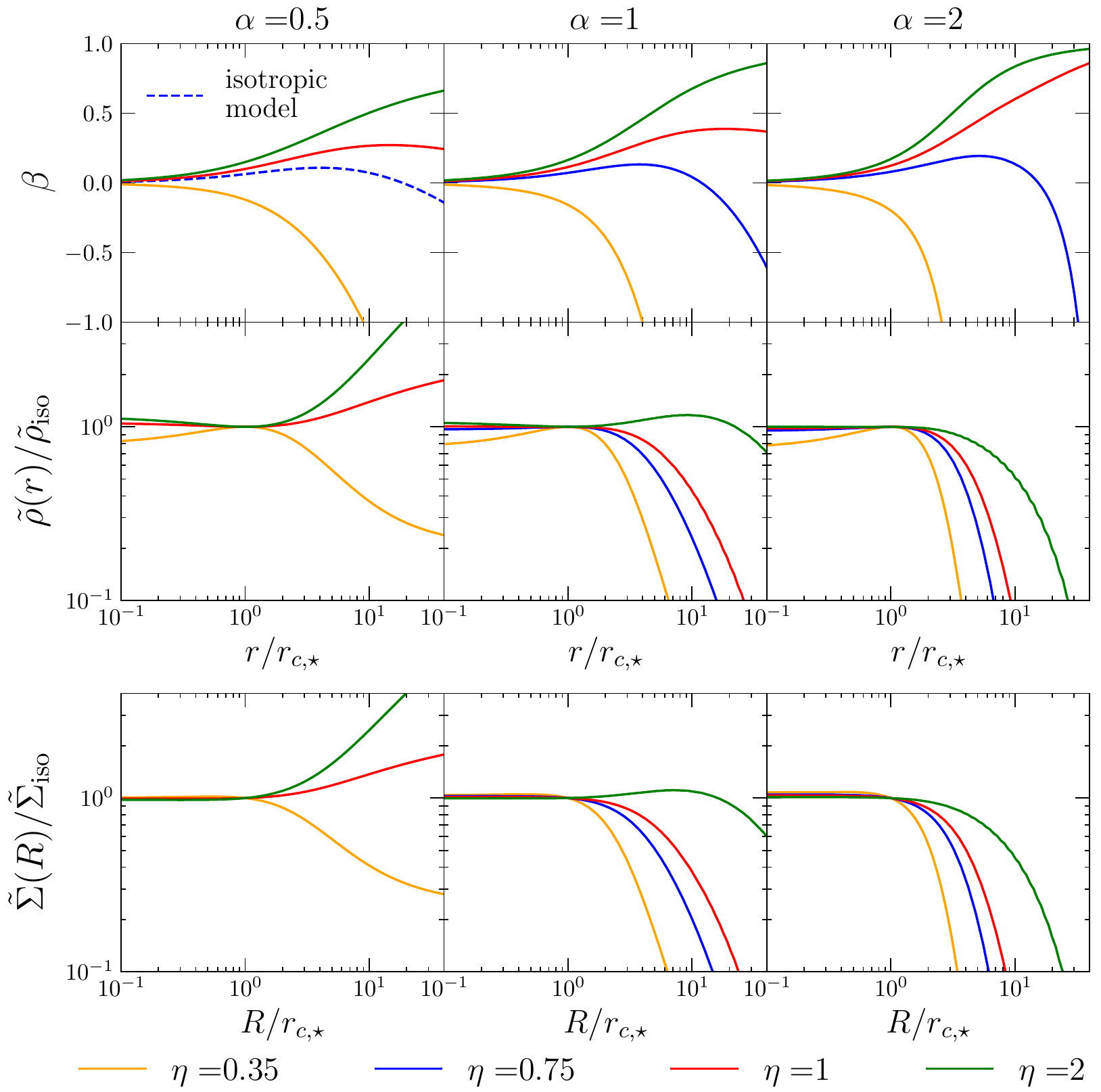}
  \caption{One-component models with, from left to right, $\alpha$ = 0.5, 1, 2. Orange, 
  blue, red and green curves refer to models with $\eta=0.35, 0.75, 1$ and $2$, respectively. 
  Top: anisotropy parameter. Centre: ratio between model normalized 
  density and normalized density of the isotropic model (Fig.~\ref{fig:isotropic}).
  Bottom: 
  same as centre row, but for surface density. In the left column we show the isotropic
  reference model $(\alpha,\eta)=(0.5,0.75)$ only the top panel (dashed blue curve). 
  Distances are normalized to the core radius $\rc$. We define 
  $\widetilde{\rho}(r) \equiv\rho(r)/\rho(\rc)$ and $\widetilde{\Sigma}(R)\equiv \Sigma(R)/
  \Sigma(\rc)$. $\widetilde{\rho}_{\rm iso}$ and $\widetilde{\Sigma}_{\rm iso}$ are the
  density and surface density profiles of the isotropic model (Fig.\ref{fig:isotropic}).
  }\label{fig:alphafix}
 \end{minipage}
\end{figure*}

\begin{figure*}
 \begin{minipage}{0.65\linewidth}
\includegraphics[width=1.\hsize]{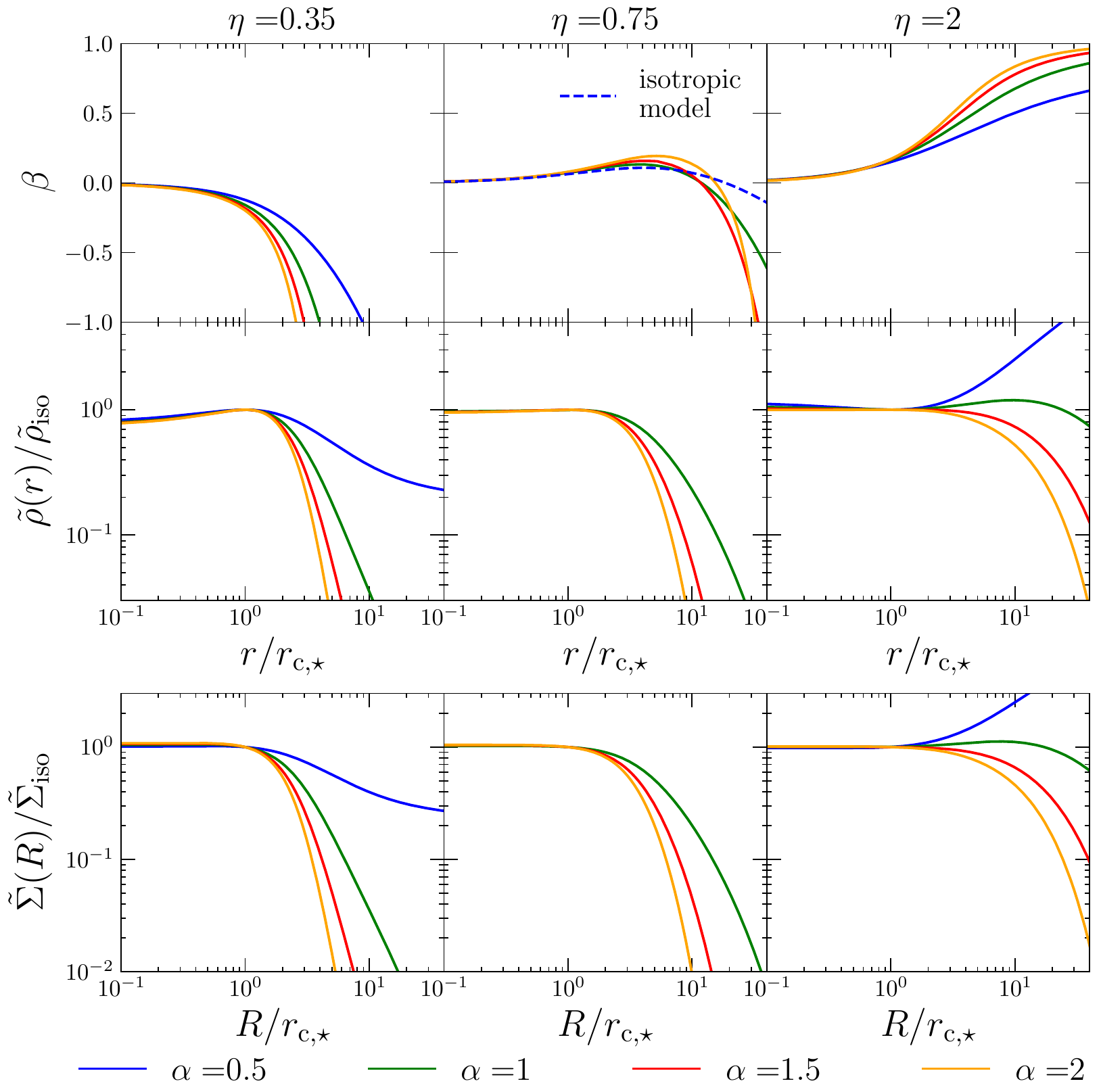}
  \caption{Same as Fig.~\ref{fig:alphafix} but now in each column $\eta$ is
  fixed to $\eta=0.35, 0.75$ and 2, from left to right. Blue, green, red and orange
  curves mark models with $\alpha=0.5, 1, 1.5$ and $2$, respectively. In the second column, 
  we show the isotropic reference model ($\alpha=0.5,\eta=0.75$) only in the top panel
  (dashed blue curve). The definitions of $\widetilde{\rho}(r)$, $\widetilde{\Sigma}(R)$ and
  $\rc$ are as in Fig.~\ref{fig:alphafix}.}\label{fig:etafix} 
 \end{minipage}
 \hfill
 \begin{minipage}{0.33\linewidth}
 \vspace{1.2cm}
  \includegraphics[width=1\hsize]{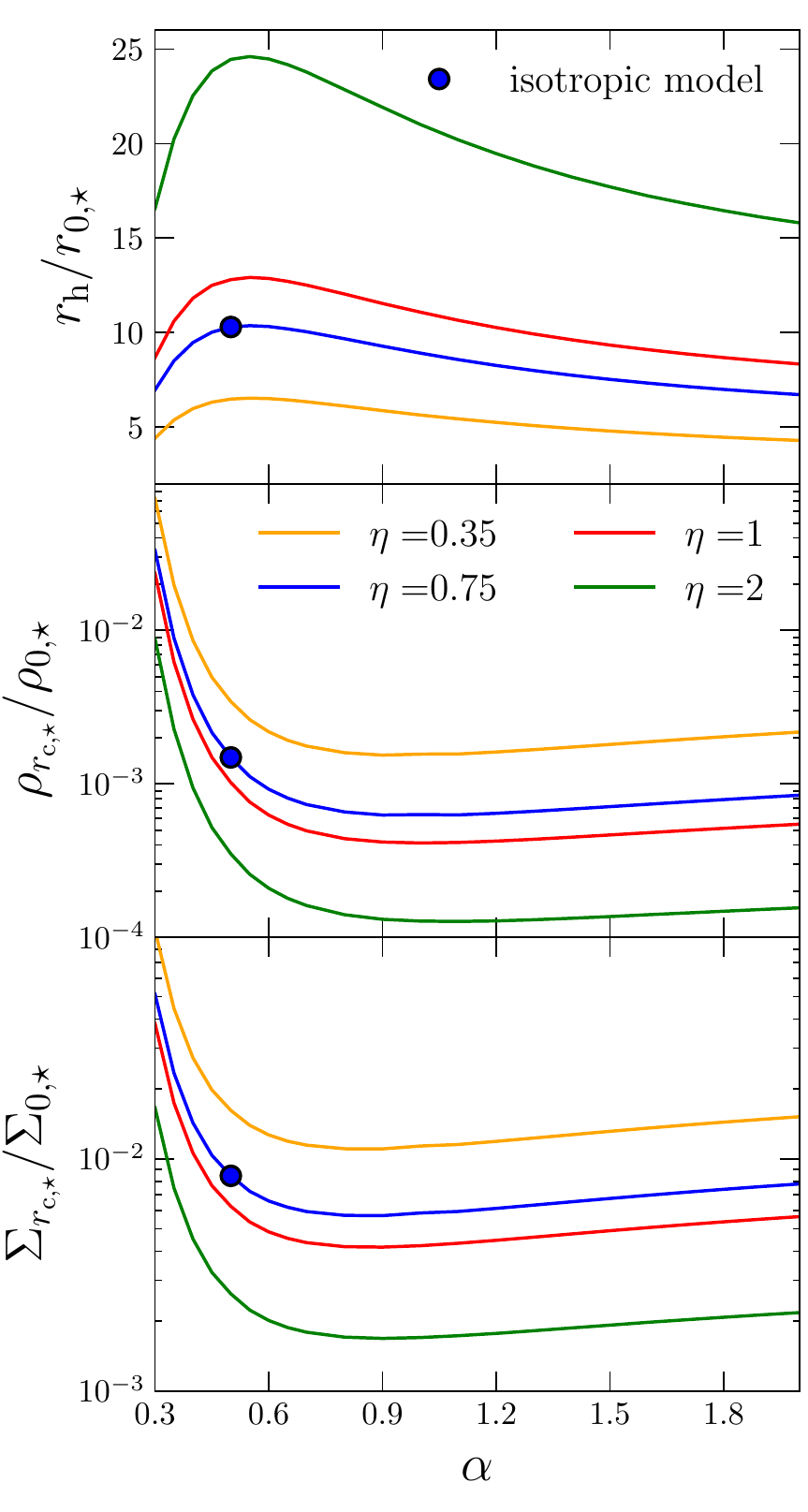}
  \caption{Measures of concentration versus $\alpha$ for one-component models with
  $\eta=0.35,\,0.75,\,1$ and $2$. 
  Top: $\rh/\rst$. Centre: $\rho_{\rc}/\rho_{0,\star}$. Bottom:
  $\Sigma_{\rc}/\Sigma_{0,\star}$.  Orange, blue, red and green curves refer 
  to models with $\eta=0.35,\,0.75,\,1$ and $2$, respectively. $\rst$ is defined by 
  equation (\ref{for:rst}), $\rho_{0,\star}\equiv\Mst/\rst^3$,
  $\Sigma_{0,\star}\equiv\Mst/\rst^2$, $\rho_{\rc}\equiv\rho(\rc)$ 
  and $\Sigma_{\rc}\equiv\Sigma(\rc)$. The blue circle marks the position 
  of the isotropic model.}\label{fig:params} 
 \end{minipage}
\end{figure*}

We define $\Jr$, $\Jphi$, and $\Jz$ as the radial, azimuthal
and vertical actions, respectively, and, following \cite{Pascale2018}, use the DF 
\begin{equation}\label{for:df1}
  \fJst = \fst\exp\biggl[-\biggl(\frac{\kJ}{\Jst}\biggr)^{\alpha}\biggr],
\end{equation}
with
\begin{equation}\label{for:omo}
 \kJ\equiv \Jr + \etaphi|\Jphi| + \etaz\Jz.
\end{equation}
The factor
\begin{equation}
 \fst = \frac{\etaphi\etaz\alpha}{(2\pi\Jst)^3 \Gamma(3/\alpha)},
\end{equation}
where $\Gamma$ is the gamma function, normalizes $\fJst$
(equation~\ref{for:unity}).  This
DF produces potentially anisotropic components with density distributions
that have cores and at large radii can be truncated in an adjustable
way.

We restrict to spherical models by fixing $\etaz=
\etaphi\equiv\eta$. In a spherical potential, $\Jphi$
and $\Jz$ are related to the total angular momentum $L$ by $L\equiv 
|\Jphi| + \Jz$, so equation (\ref{for:omo}) reduces to 
\begin{equation}\label{for:omo2}
 \kJ = \Jr + \eta(|\Jphi| + \Jz) = \Jr + \eta L.
\end{equation}

We define the stellar core radius $\rc$ as the radius where 
\begin{equation}
\gamma_{\star}\equiv \frac{\text{d}\ln\rhost}{\text{d}\ln r}=-\fracj{1}{2}. 
\end{equation}
We define the 
half-mass radius $\rh$ to be the radius of the sphere containing half of the
stellar mass, and the effective radius $\Reff$ to be the radius on the
plane of the sky that contains half of the projected mass.

With $\sigma_t$ and $\sigma_r$ the velocity dispersions in the 
tangential and radial directions, respectively,
\begin{equation}\label{for:beta}
 \beta \equiv 1 - \frac{\sigma_t^2}{2\sigma_r^2}
\end{equation}
measures the amount of velocity anisotropy. Isotropic velocity distributions 
correspond to $\beta = 0$, tangentially biased ones to $\beta < 0$ and radially 
biased ones to $0 < \beta \le 1$. 

We briefly comment on the physical meaning of the relevant free parameters
of the DF (\ref{for:df1}) when the latter is multiplied by the stellar mass 
$\Mst$.

\begin{itemize}

 \item $\Jst$: the action scale that naturally defines the length scale
 \begin{equation}\label{for:rst}
  \rst =  \frac{\Jst^2}{G\Mst}
 \end{equation}
 and the velocity scale
 \begin{equation}\label{for:vst}
  \vst = \frac{G\Mst}{\Jst}.
 \end{equation}
 Any pair among $\Mst$, $\Jst$, $\rst$ and $\vst$, sets the model's physical 
 scales and can be adjusted to match some physical property of a target 
 system (for instance, the total mass or the central velocity dispersion).
 
 \item $\alpha$: a non-negative, dimensionless parameter that mainly
 regulates the model's density profile.
 
 \item $\eta$: a non-negative, dimensionless parameter that mainly controls 
 the radial or tangential bias velocities of the model; models sharing 
 the parameters ($\alpha$, $\eta$) are homologous.
\end{itemize}

In the case of spherical symmetry ($\eta_\phi=\eta_z$), the DF (\ref{for:df1})
can be considered a generalization of the spherical anisotropic Michie-King
DF $f(E,L)$. 
Dealing with actions $\vJ$ rather than $(E,L)$ facilitates extension to 
multi-component and flattened  models (\citealt{Binney2014}). Models generated 
by the DF (\ref{for:df1}) lack rotation, but the model can be set rotating without
changing the density distribution by adding a DF that is odd in $\Jphi = \Lz$.

\subsection{Spherical one-component models}
\label{subsec:oneC}

Fig.~\ref{fig:isotropic} plots the general properties of a nearly-isotropic model, 
obtained with $(\alpha,\eta)=(0.5,0.75)$. Panel a shows that the model is almost
isotropic along the whole radial extent, with $|\beta|\le0.1$ out to $r\simeq30\rc$. 
Panel c shows that the density distribution is cored, so $\gamma_{\star}\simeq0$ 
near the centre, and is exponentially truncated farther out, so $\gamma_{\star}
\lesssim-3$ at $r\simeq\rh$ (panel b). The fact that an almost isotropic model 
is obtained when $\eta=0.75$ can be explained as follows. Since the DF $\fiso$ 
of an isotropic model can depend on only the Hamiltonian $H$, it will satisfy
\begin{equation}\label{for:freq}
\frac{\partial \fiso}{\partial L}\biggr/\frac{\partial \fiso}{\partial \Jr} =
\frac{\Omega_L}{\Omega_r},
\end{equation}
where $\Omega_L=\DD H/\DD L$ and $\Omega_r=\DD H/\DD J_r$ are, respectively,
the tangential and radial frequencies. We expect $\Omega_L/\Omega_r$ to be a 
smooth function of ${\bf J}$, ranging from 1/2 for small actions (where $\Phi$ 
is almost simple-harmonic) to 1 for large actions (where $\Phi$ is almost
Keplerian). However, the DF (\ref{for:df1}) is such that
\begin{equation}\label{for:freqeta}
\frac{\partial \fJst}{\partial L}\biggr/\frac{\partial \fJst}{\partial \Jr} =\eta,
\end{equation}
independent of the actions. The choice $\eta\simeq0.75$ reasonably ensures 
a good compromise between the expected $\Omega_L/\Omega_r$ in the two regimes of 
small and large actions. 

Fig.s~\ref{fig:alphafix} and \ref{fig:etafix} show how $\alpha$ and $\eta$
affect a model's 
anisotropy and density profiles by comparing them with those of the reference 
isotropic model. The parameter $\eta$ mainly regulates the orbital anisotropy 
(Fig.~\ref{fig:alphafix} top row). Models are isotropic when $r\lesssim\rc$
because no model with a cored density distribution can be 
radially anisotropic inside the core (\citealt{An2006},
\citealt{Ciotti2010}). 
In the outer regions, a model can be either tangentially or 
radially biased. Anisotropy is mildly enhanced by increasing $\alpha$:  
tangentially biased models become more tangential and radially biased
models become more radial (Fig.~\ref{fig:etafix} top row).

 Let the normalized density profile be $\widetilde{\rho}
\equiv\rho/\rho(\rc)$ and the normalized surface density profile be
$\widetilde{\Sigma}\equiv\Sigma/\Sigma(\rc)$, and call these quantities for the
isotropic model $\widetilde{\rho}_{\rm iso}$ and
$\widetilde{\Sigma}_{\rm iso}$, respectively. Then the middle and bottom rows of 
Fig.s~\ref{fig:alphafix} and \ref{fig:etafix}, show, respectively, the
profiles of $\widetilde{\rho}/\widetilde{\rho}_{\rm iso}$ and $\widetilde{\Sigma}/
\widetilde{\Sigma}_{\rm iso}$. We see that $\alpha$ and $\eta$ are degenerate in determining
the density profile.  Increasing $\alpha$ truncates the DF
(\ref{for:df1}) more rapidly for large actions, while decreasing $\eta$
encourages orbits with high angular momentum. In either case, the outer
density profile steepens.  Increasing $\eta$ favours eccentric orbits and
thus makes the
density distribution slightly more cuspy (Fig.~\ref{fig:alphafix} middle
row). Conversely, very tangentially biased models may present a density
minimum at the centre (Fig.~\ref{fig:etafix} middle left panel).

One could make $\eta$ a function of $\vJ$ to achieve greater flexibility in
the anisotropy (see \citealt{Williams2015b}), but the simple choice of constant 
$\eta$ provides significant flexibility (Fig.s~\ref{fig:alphafix} and
\ref{fig:etafix}), and avoids the introduction of new free parameters. 
We find empirically that models with $\eta>2$ or $\alpha>2$ have properties 
very similar to models with $\eta=2$ or $\alpha=2$, so we do not show them here. 

Fig.~\ref{fig:params} shows how the physical scales $\rh/\rst$, 
$\rho_{\rc}/\rhos$ and $\Sigma_{\rc}/\sigmas$ vary with 
($\alpha,\eta$). Here $\rho_{\rc}\equiv\rho(\rc)$, $\Sigma_{\rc}\equiv
\Sigma(\rc)$, $\rhos\equiv\Mst/\rst^3$ and $\sigmas\equiv\Mst/\rst^2$. 
When $\eta$ is  decreased at fixed $\alpha$, the model becomes more compact
(middle row of  Fig.~\ref{fig:alphafix}), so $\rh/\rst$
decreases and $\rho/\rhos$ increases.
While changing $\alpha$ at fixing $\eta$ affects the physical scaling 
only when $\alpha\lesssim0.5$: in this regime, $\rh/\rst$ shortens 
(Fig.~\ref{fig:params}a) and models are slightly more cuspy, moving
$\rho_{\rc}$ to higher values.

\begin{figure*}
 \centering
 \includegraphics[width=1\hsize]{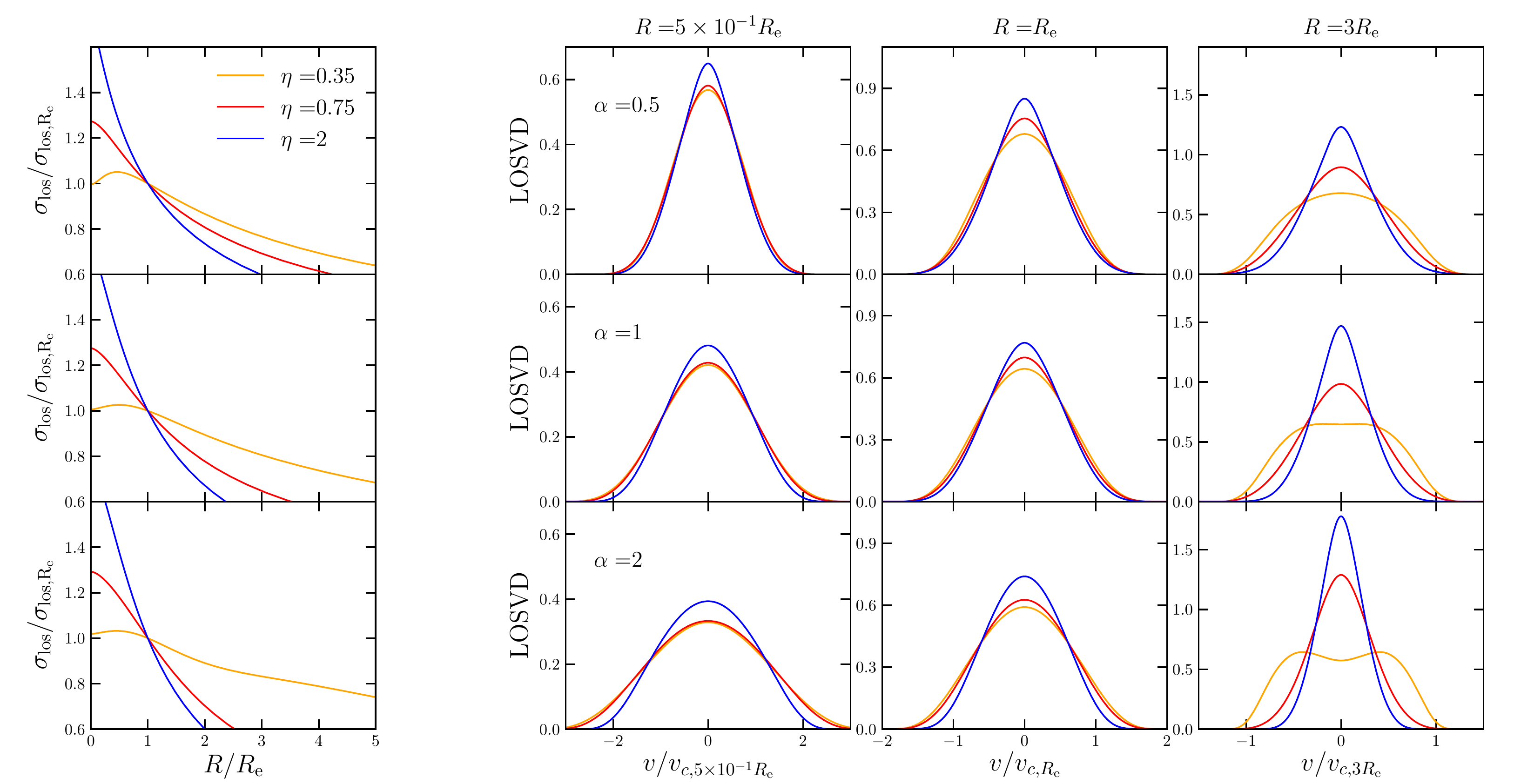}
 \caption{Kinematic observables in one-component models.  Orange, red and
blue curves are for models with to $\eta=0.35, 0.75$ and $2$, respectively. Panels
in the left column show line-of-sight velocity dispersions, normalized to 
$\sigma_{\rm los, \Reff}\equiv\sigma_{\rm los}(\Reff)$, when $\alpha=0.5, 1, 2$.
The other three panels show LOSVDs with the velocity scale normalized to $v_{c, R}$,
the circular speed at the radius of observation $R$, which increases from left to
right: $R=5\times10^{-1}\Reff$, $R = \Reff$ and $R=3\Reff$. The value of
$\alpha$ increases from top to bottom: $\alpha=0.5,1,2$.}\label{fig:losvdoneC}
\end{figure*}

Fig.~\ref{fig:losvdoneC}  plots the line-of-sight velocity dispersion profiles
of models with different values of $\alpha$ and $\eta$, together with LOSVDs 
at three radii. The shape of a LOSVD encodes the velocity anisotropy: a
flat-topped LOSVD indicates a tangentially biased system, while a radially
biased system yields peaky LOSVDs. A wide LOSVD reflects highly populated
nearly circular orbits: note how models with $\eta=0.35$ generate the widest
LOSVDs. The model with $(\alpha, \eta) = (2,0.35)$ is an example of a model
with extreme tangential anisotropy, in which the LOSVD is double-peaked
around plus and minus the circular speed. This generates the flattest
line-of-sight velocity dispersion profiles (Fig.~\ref{fig:losvdoneC}, left
column).


\subsection{Spherical two-component models}
\label{subsec:twoC}

We focus now on two-component spherical models, consisting
of a stellar population with DF (\ref{for:df1}) and a dark halo 
with DF $\fJdm$. 
The adiabatic invariance of the actions makes them natural tools with which
to analyse the addition of a stellar component to a dark halo. In
the simplest scenario, gas falls into a dark halo over many dynamical times,
so the dark halo contracts adiabatically. In this case the dark halo's
present configuration can be computed from its original DF $f(\vJ)$ without knowing
how the rate of accretion of baryons varied over cosmic time. The dark halo
is then predicted to have a very cuspy central structure, 
comprising particles with very small velocities. So, it would not be
surprising if fluctuations in the gravitational potential generated by the
baryons before most of them were driven out by supernovae had
upscattered the least energetic dark-matter particles and thus erased the
cusp \citep{NavarroEF1996,Governato2012, NipotiBinney2015,ReadWS2019}.
For this reason we explore models with a dark-matter DF 
that, depending on the value of a parameter $\Jcdm$, generates 
either a classical cuspy halo or a cored halo. This DF is \citep{ColeBinney2017,Pascale2018}
\begin{equation}\label{for:dmdf}
 \fJdm = \gcJ\gNFW\TJ,
\end{equation}
where 
\begin{equation}\label{for:CBdf}
 \gcJ = \biggl[\biggl(\frac{\Jcdm}{\hJ}\biggr)^2 -\mu\frac{\Jcdm}{\hJ} + 1\biggr]^{-5/6}, 
\end{equation}
\begin{equation}\label{for:PostiDF}
 \gNFW = \frac{\gdm}{\Jdm^3}\frac{[1 + \Jdm/\hJ]^{5/3}}{[1 + \hJ/\Jdm]^{2.9}}
\end{equation}
and
\begin{equation}\label{for:PasDF}
 \TJ = \exp{\biggl[-\biggl(\frac{\hJ}{\Jtdm}\biggr)^{2}\biggr]}.
\end{equation}
Here $\hJ$ is a homogeneous function of the actions of degree one
\begin{equation}
 \hJ = \Jr + \hphi|\Jphi| + \hz\Jz.
\end{equation}
 The core action $\Jcdm$ sets the spatial
extent of the core in the density distribution, while $\mu$ is a dimensionless
parameter used to make the dark-matter mass (\ref{for:dmdf}) independent of
$\Jcdm$ (\citealt{ColeBinney2017}).  This convention is motivated by the idea that
non-zero $\Jcdm$ arises through dark-matter particles being upscattered but
not ejected from the halo. $\Jtdm$ is the truncation action, which
serves to make normalization of the DF possible (equation~\ref{for:unity}).

We set the dimensionless parameters $\hphi$ and $\hz$ to a common value $h$
so the DF (\ref{for:dmdf}) generates spherical models.  In this case we
cannot give an analytic expression for the constant $\gdm$, which normalizes
$\fJdm$ to unity. However, it can be readily computed following Appendix A of
\cite{Pascale2018}.  The total dark matter mass $\Mdm$ together with the
action scale $\Jdm$ define via equations (\ref{for:rst}) and (\ref{for:vst}) a
scale radius $\rdm$ and a scale velocity $\vdm$.

\cite{Posti2015} showed that, in isolation, the DF (\ref{for:PostiDF}) generates 
NFW-like models. The factor (\ref{for:CBdf}) was added by \cite{ColeBinney2017}
to enable the DF to describe cored NFW models.\footnote{The DF (\ref{for:PostiDF})
is singular for $||{\bf J}||\to0$, and equation (\ref{for:CBdf})
compensates for such divergence, making the central phase-space density
finite.} 
The factor (\ref{for:PasDF}) was introduced by \cite{Pascale2018}.

\begin{table}
 \begin{center}
  \caption{Scale radii and corresponding circular speeds for one-component
  NFW haloes expressed in terms of the characteristic radius and velocity
  that follow from the halo's DF (equations \ref{for:rst} and \ref{for:vst} with
  $\star$
  replaced by dm). Equation (\ref{for:dmrs}) defines $\rsdm$.}
  \begin{tabular}{cc}
  \hline\hline
  $\rsdm/\rdm$	&	$\vcdm(\rsdm)/\vdm$	\\
  0.67		&	0.40		\\
  \hline\hline
  \end{tabular}
  \label{tab:NFWoneC}
 \end{center}
\end{table}

As in \cite{Pascale2018}, we define the dimensionless parameters 
\begin{equation}\label{for:Jc}
 \Jctil \equiv \Jcdm/\Jdm, 
\end{equation}
\begin{equation}\label{for:Jdm}
\Jdmtil\equiv \Jdm/\Jst. 
\end{equation}
\begin{equation}\label{for:Jt}
 \Jttil = \Jtdm/\Jdm, 
\end{equation}
and
\begin{equation}\label{for:Mdm}
\Mdmtil \equiv \Mdm/\Mst.
\end{equation}
Models sharing $\alpha$, $\eta$, $\Jctil$, $\Jdmtil$, $\Jttil$, $\Mdmtil$, $\mu$  and $h$ 
are homologous. The physical scales can be set a posteriori by choosing any pair among 
$\Mdm, \Jdm$, $\rdm$ and $\vdm$.
We introduce the logarithmic 
slope of the dark-matter density $\gammadm\equiv\text{d}\ln\rhodm/\text{d}\ln r$,
and define the halo scale radius $\rsdm$ from the relation 
\begin{equation}\label{for:dmrs}
\gammadm(\rsdm) = -2,
\end{equation}
as for the classical NFW model. The truncation and core radii are defined by
\begin{equation}
\gammadm(\rtdm) = -3,
\end{equation}
and
\begin{equation}
 \gammadm(\rcdm) = -\fracj{1}{2},
\end{equation}
respectively.

\subsubsection{Impact of stars on dark haloes}
\label{subsec:contraction}

We consider representative stellar components with several orbital
anisotropies, and examine the effects that cuspy or cored dark haloes and
stars have on each other when they cohabit in the potential they jointly
generate.  we set $\alpha=0.5$ and select stellar DFs
(\ref{for:df1}) that generate, in isolation, tangential, isotropic and
radially biased models, by fixing $\eta=0.35 ,0.75$ and $1$, respectively
(Section~\ref{subsec:oneC}).  For fixed $\Mdm$, we
vary $\Mst$ to control the relative mass contribution $\Mdmtil$.  For both
cuspy and cored haloes, and for each stellar anisotropy, we consider three
groups of models, with $\Mdmtil=10^4, 10^3, 10^2$. We refer to them as
DM$_i$-NFW for the NFW haloes, and DM$_i$-Cored, for the cored haloes, with
$i=1,2,3$, respectively.  As $\Mdmtil$ decreases, the stellar component
becomes more massive.  The chosen values of $\Mdmtil$ generate models in which the dark
halo strongly dominates over the stars in the central parts (DM$_1$,
$\Mdmtil(\rh)\gtrsim20$), models in which stars and dark matter have similar
density in the central parts (DM$_2$, $\Mdmtil(\rh)\simeq1$), and models in
which the stars dominate in the central parts (DM$_3$,
$\Mdmtil(\rh)\lesssim0.1$). In all groups the dark matter dominates far out.
We do not explore different dark-halo anisotropies (for details, see
\citealt{PifflPenoyreBinney2015}) but set $h=1$, which makes the dark halo
slightly radially biased.  Also, $\Jdmtil=3000$, which ensures
$\rsdm/\rh>1$ in all cases.

The exponential cut-off (\ref{for:PasDF}) introduces much freedom in setting
$\Jttil$, which, as long as it is large enough, does not affect the halo's
central properties. Thus, we standardize on $\Jttil=20$, which truncates the
halo density sufficiently far from the scale radius that it has no impact in
the observationally accessible region ($\rtdm/\rsdm\gtrsim30$).  Once
$\Jttil$ and $h$ have been set for an NFW model, the DF's physical scales
follow unambiguously --
Table~\ref{tab:NFWoneC} lists the values of $\rsdm/\rdm$ and $\vcdm(\rsdm)/\vdm$
(i.e. the halo circular speed computed at $\rsdm$). The quantities $\rsdm$
and $\vcdm(\rsdm)$ are available from cosmological simulations, and the pair
($\rdm$, $\vdm$) can be easily computed from Table~\ref{tab:NFWoneC} to
scale any $\fJ$ NFW-model onto the required scales.  For the cored
models we chose $\Jctil=0.02$, which implies $\mu=0.2117$. The resulting core
radius is $\rcdm\simeq0.1\rsdm$. Table~\ref{tab:twotab} summarizes the
relevant parameters used to generate the presented models. 

\begin{figure*}
 \centering
 \includegraphics[width=1.\hsize]{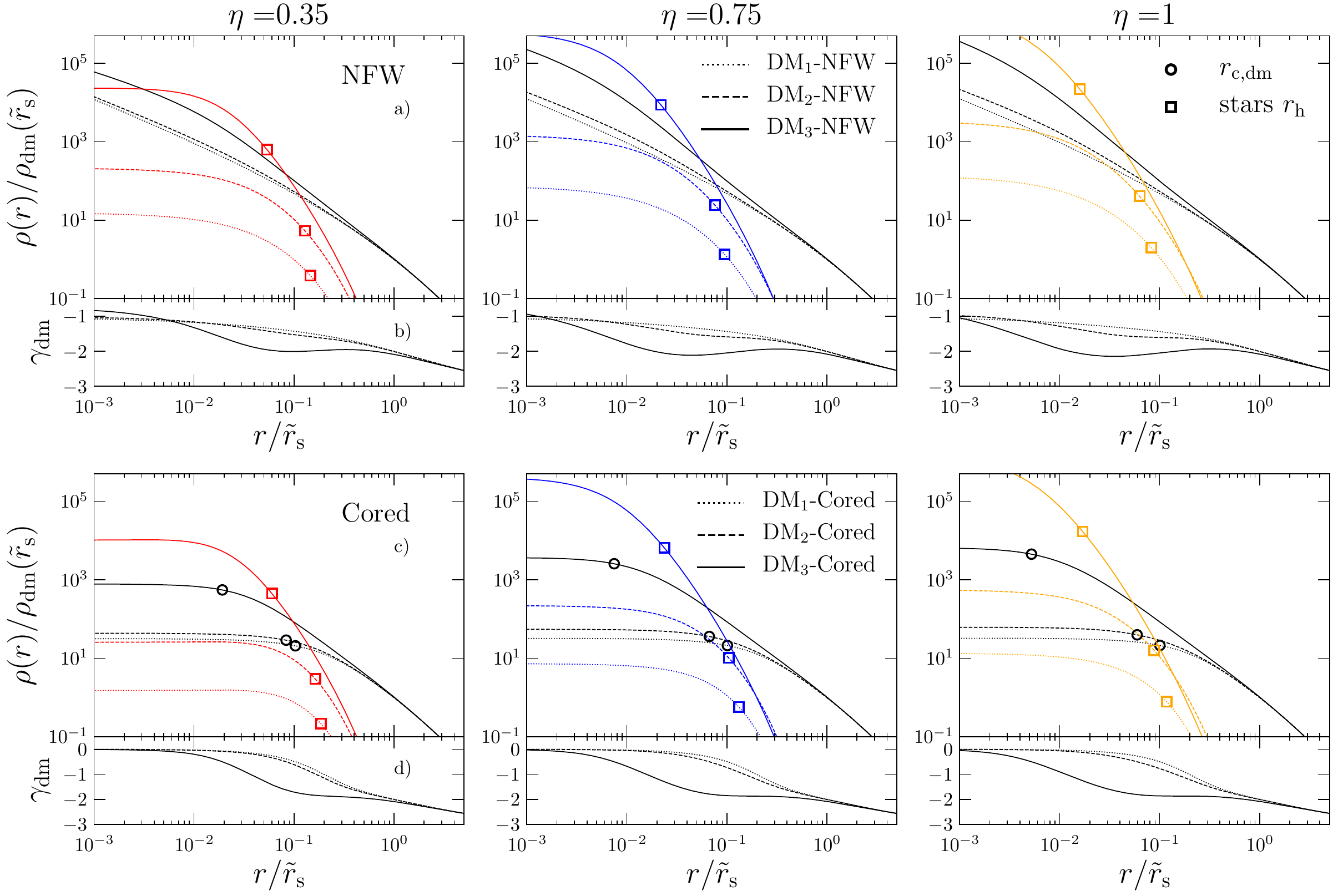}
 \caption{Density profiles of models with both stellar and dark-matter
 components with radii normalized to the scale radius $\widetilde{r}_{\rm s}$
 of the corresponding one-component halo. Models shown in the upper row 
 have NFW haloes, while models shown in the lower row have cored haloes. 
 Table~\ref{tab:twotab} lists the models' parameters. 
 The stellar mass fraction increases along the sequence dotted, dashed,
 full curves. DM profiles are plotted in black while stellar profiles are
 coloured. Squares indicate the half-mass radii of the
 stellar components, while in the lower row the 
 black circles mark the DM core radii, defined by $\gammadm(\rcdm)=-1/2$. 
 }\label{fig:contraction}
\end{figure*}

Fig.~\ref{fig:contraction} plots, for our two-component models, the
profiles of the dark matter (black curves) and stars (coloured curves), and
also the dark-matter logarithmic density slopes $\gammadm$ (long-thin bottom
panels). Models with NFW haloes are plotted in the top row, while the bottom
row shows models with cored haloes. The left column shows models with
tangentially biased stellar components, while the rightmost column shows
models in which the stellar component is radially biased. 
Dotted ($i=1$), dashed ($i=2$) and full ($i=3$) black lines show the
dark haloes of models with increasingly massive stellar components. Whereas
the dark haloes differ only modestly between $i=1$ and $i=2$, once the case
$i=3,\Mdmtil=10^2$ is reached, the stars' gravity enhances the central density
of the halo by a factor $\sim10$ in the case of an NFW halo, and by a larger
factor in the case of a cored halo. In all the $i=3$ models, the
halo-steepness parameter hangs around $-2$ over a wide range of radii
interior to $\rsdm$ with the consequence that the scale radius $\rsdm$ of
these models is not uniquely defined.  The steepening of $\gammadm$ can
reduce the core radii $\rcdm$ of cored models by a factor 10.  The ratio
$\rcdm/\rh$ also reduces, by a factor $\sim2$. This reduction diminishes the
extent of the stellar system that is dominated by the halo's core. Radially
biased stellar components contract their  dark haloes more strongly than
tangential biased  ones because radial bias increases the central star
density (Fig.s~\ref{fig:alphafix}, \ref{fig:etafix}, and \ref{fig:params}).

\subsubsection{Impact of dark haloes on stars} 
\label{subsec:NFW}

\begin{table}
 \begin{center}
  \caption{Parameters used to generate the representative,
  two-component models: $h$, dimensionless parameter regulating 
  the anisotropy of the dark halo; $\Jdmtil$ and $\Jttil$ as in equations 
  (\ref{for:Jdm}) and (\ref{for:Jt}); $\alpha$ and $\eta$, defined by the DF 
  (\ref{for:df1}); $\Mdmtil$ as in (\ref{for:Mdm}); $\Jctil$, as in equation 
  (\ref{for:Jc}); $\mu$, dimensionless parameter used to make the normalization
  of the DF (\ref{for:dmdf}) independent of $\Jcdm$.}  \begin{tabular}{cccccc}
  \hline\hline
  $h$	&	$\Jdmtil$ 	&	$\Jttil$	&	$\alpha$ &	$\Mdmtil$	&	$\eta$		\\
  \hline\hline
  1	&	3000		&	20		&	0.5	 &$10^3$-$10^4$-$10^5$	&	$0.35$-$0.75$-$1$\\
  \hline\hline
  \multicolumn{4}{c}{NFW models}			& 	\multicolumn{2}{c}{Cored models}			\\
  \multicolumn{4}{c}{($\Jctil,\mu$)=(0,0)}		&	\multicolumn{2}{c}{($\Jctil, \mu$)=(0.02,0.2117)}	\\
  \hline\hline
  \end{tabular}
\label{tab:twotab}
 \end{center}
\end{table}

\begin{figure*}
 \centering
 \includegraphics[width=1.\hsize]{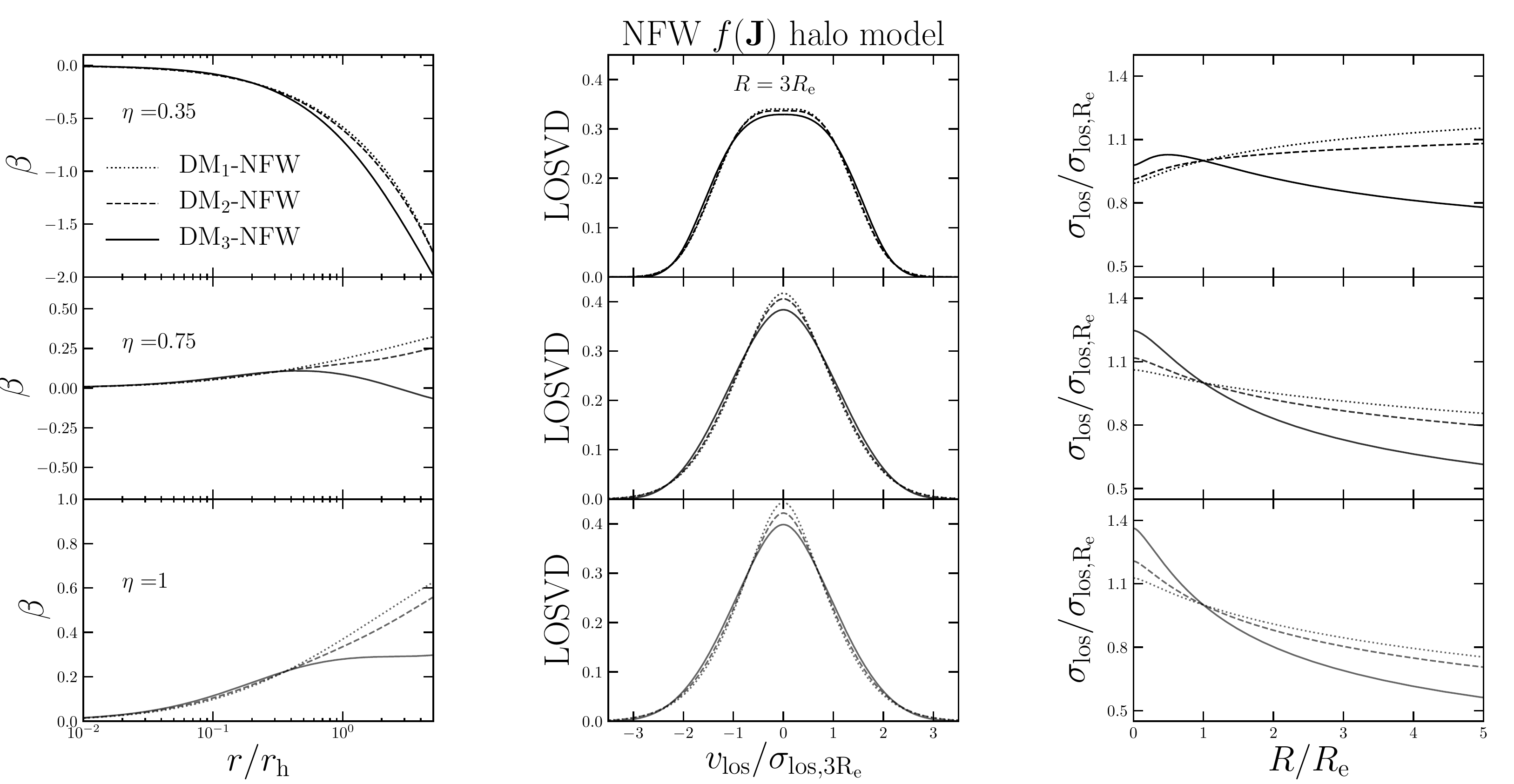}
 \caption{Stellar kinematics in two-component models with NFW
 dark-matter haloes (equation \ref{for:dmdf}). The stellar mass fraction
 increases along the sequence dotted, dashed, full curves. All models have
 $\alpha=0.5$ while $\eta$ increases from top to bottom ($\eta=0.35,0.75,1$)
 so the top and bottom models are tangentially and radially  biased,
 respectively. The left column
 shows the anisotropy parameter, the centre column shows the LOSVD at
 $R=3\Reff$ normalized to the local velocity dispersion, and the right column shows $\sigma_{\rm los}$ normalized to its
 value at $\Reff$. In all models $\Jdmtil=3000$, $\Jttil=20$, $\Jctil=0$, $\mu=0$, $h=1$,
 $\alpha=0.5$ (Table~\ref{tab:twotab}).
 }\label{fig:starNFW}
\end{figure*}

\begin{figure*}
 \centering
 \includegraphics[width=1.\hsize]{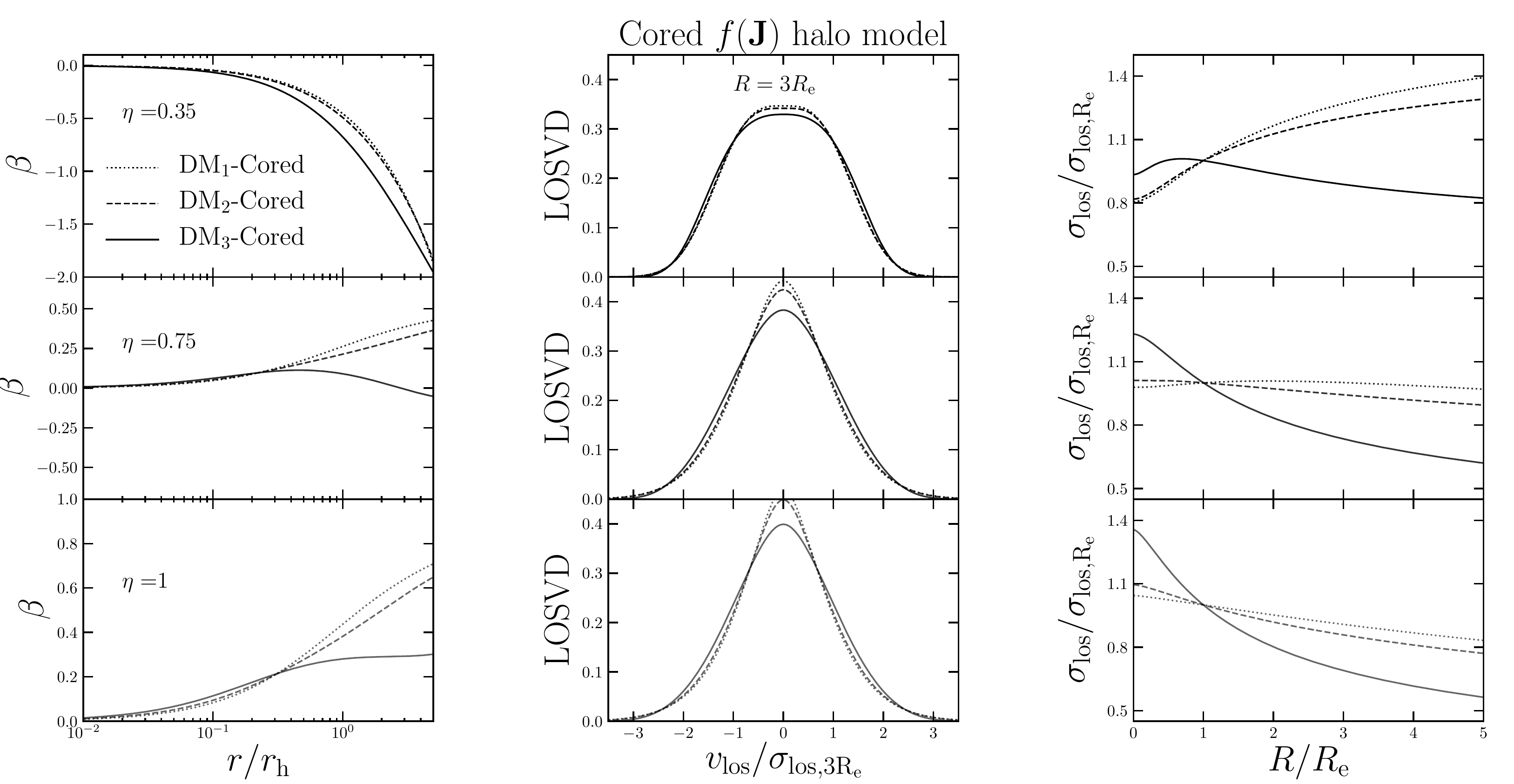}
 \caption{As Fig.~\ref{fig:starNFW} except for models with cored haloes 
 (see Table~\ref{tab:twotab}).}\label{fig:starcore}
\end{figure*}

Fig.s~\ref{fig:starNFW} and \ref{fig:starcore} show, respectively, the impact NFW and cored
haloes have on the kinematics of
the stellar component. Again dotted, dashed and full curves relate to
increasingly massive stellar components  ($i=1,2,3$), and black, grey and light grey curves
relate to tangentially biased, isotropic and radially biased stellar
components. Addition of a dark halo changes the velocity anisotropy of the
stellar component (left column) by decreasing the ratio $\Omega_L/\Omega_r$
at a given radius, and it is this ratio which sets the value of $\eta$ that
corresponds to 
isotropy (equations~\ref{for:freq} and \ref{for:freqeta}). Since adding a halo
diminishes the critical value of $\Omega_L/\Omega_r$, at fixed $\eta$ it 
increases radial bias (broken curves above full curves in left columns of 
Fig.s~\ref{fig:starNFW} and \ref{fig:starcore}). This effect is most
pronounced at $r\gg\rh$, where the potential of a one-component model is 
almost Keplerian.

These changes in $\beta$ make the LOSVD at $R=3\Reff$, shown in the central
columns, more peaky, but the effect is quite weak and would be very hard to
detect observationally. The right columns plot $\sigma_{\rm los}(R)$, which
is significantly flattened by the addition of a massive dark halo, a
consequence of adiabatic compression of the envelope of the stellar system by
the very extended dark-matter distribution.

\subsection{Effects of a central massive black hole} 
\label{subsec:BH}
Here we explore how stellar components with the DF (\ref{for:df1}) are modified 
by a central massive black hole (hereafter BH). We present models
with and without a dominating dark halo. The potential of the BH is
taken to be that of a Plummer model
\begin{equation}\label{for:phiBH}
 \PhiBH = -\frac{G\MBH}{\sqrt{r^2+a^2}},
\end{equation}
with $a$ too small to impact any observable. 

We choose two representative stellar components with  $\alpha=0.5$ that, in isolation, are 
quasi-isotropic ($\eta=0.75$) and radially biased ($\eta=1$). 
When a dark halo is included, its parameters  are those of the DM$_2$-NFW model  
(Section~\ref{subsec:contraction}, Table~\ref{tab:twotab}). 
We consider  BH masses of $\muBH\equiv\MBH/\Mst\equiv0.001,0.0017,0.005$ 
(\citealt{Magorrian1998}). The BH's radius of influence $\Rinf$ is the projected 
distance on the plane of the sky within which the BH's gravity cannot be neglected.
We define it such that (\citealt{BinneyTremaine2008}) 
\begin{equation}\label{for:Rinf}
 \sigma_{\rm los, \star}(\Rinf) = \sqrt{\frac{G\MBH}{\Rinf}},
\end{equation}
 where $\sigma_{\rm los, \star}$ is the stars line-of-sight velocity
dispersion. Table~\ref{tab:BHtab} lists the parameters of
our models, including $\Rinf$.

\begin{table}
 \begin{center}
  \caption{Main parameters of models with BHs. The stellar  DF has the form
  (\ref{for:df1}) and the  
  BH's potential is given by (\ref{for:phiBH}). If the model has
  dark halo, its (NFW) DF is given by (\ref{for:dmdf}).  All models have
  $\alpha=0.5$ and  $\eta$
  controls the stellar anisotropy. The
  BH-to-stellar mass fraction is $\muBH=\MBH/\Mst$. Equation
  (\ref{for:Rinf}) defines the radius of influence $\Rinf$, which is given as
  a fraction of the stellar effective radius. For 
  models with dark matter, the quantities defined by equations
  (\ref{for:Jt}), (\ref{for:Mdm}) and (\ref{for:Jdm}) are $\Jttil=20$,
  $\Mdmtil=1000$, $\Jdmtil=3000$. $\Mst$ 
  and $\Jst$ can be scaled to any values of interest.}\label{tab:BHtab}
  \begin{tabular}{llcc}
  \hline\hline
		&		&		no DM		&	with DM		\\
    $\eta$	&	$\muBH$	&	$\Rinf/\Reff $ 		&	$\Rinf/\Reff $	\\
  \hline\hline
		&	0.005	&	$2.81\times10^{-2}$	&	$9.67\times10^{-3}$	\\
      0.75	&	0.0017	&	$9.87\times10^{-3}$	&	$3.31\times10^{-3}$	\\
		&	0.001	&	$6.02\times10^{-3}$	&	$2.00\times10^{-3}$	\\
  \hline
      		&	0.005	&	$2.43\times10^{-2}$	&	$9.94\times10^{-3}$	\\
      1		&	0.0017	&	$8.60\times10^{-3}$	&	$3.40\times10^{-3}$	\\
      		&	0.001	&	$5.24\times10^{-3}$	&	$2.05\times10^{-3}$	\\
  \hline\hline
  \end{tabular}
 \end{center}
\end{table}

\begin{figure*}
 \includegraphics[width=.7\hsize]{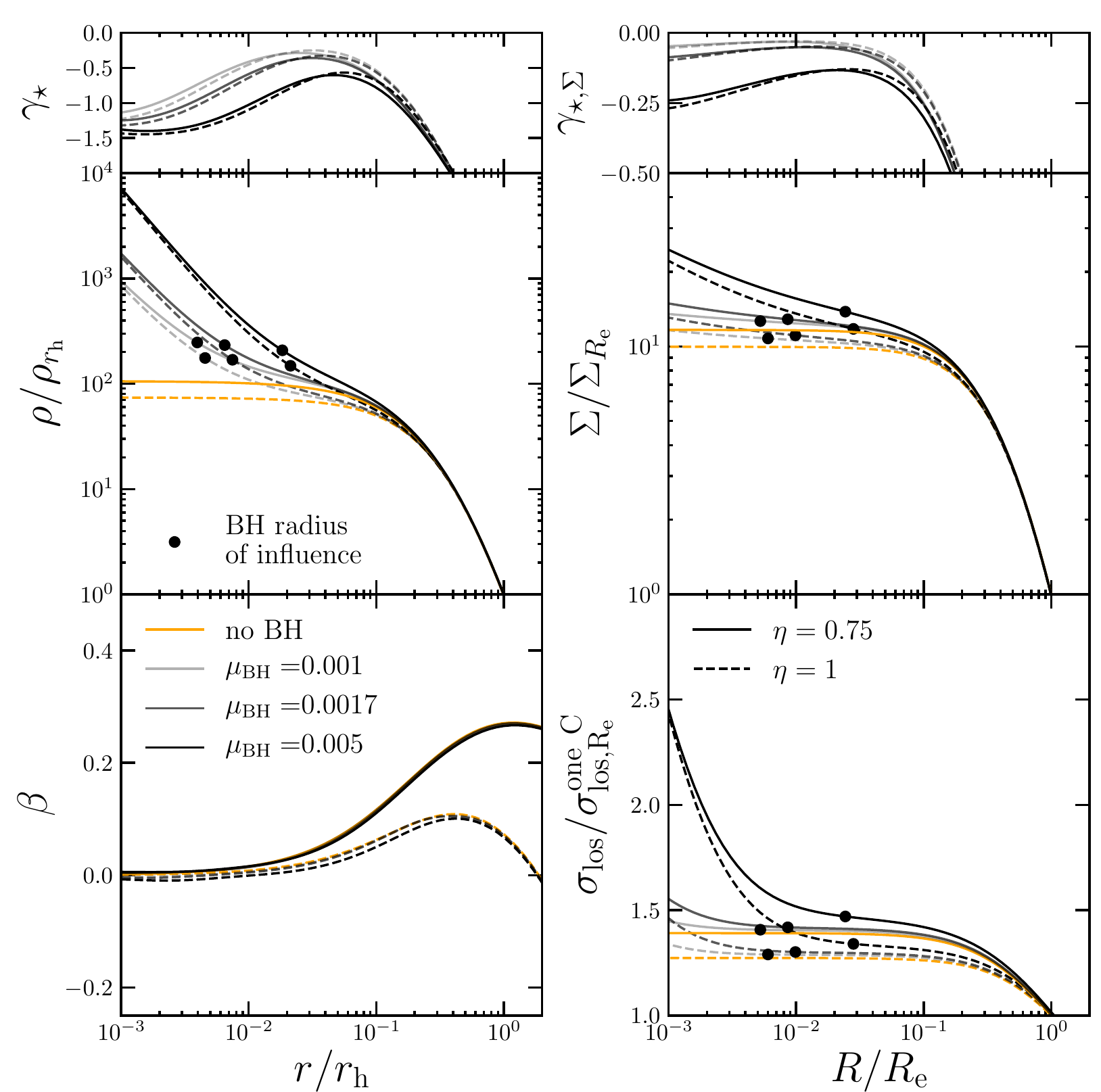}
 \caption{Impact of a central black hole of mass $\muBH\equiv \MBH/\Mst$ on
 two one-component models. A model with $\eta=0.75$ is shown by full curves 
 while dashed curves show the radially biased model with $\eta=1$. Both 
 models have $\alpha=0.5$. Colours indicate BH mass fraction: orange for 
 no black hole, greys for $\muBH=0.001,0.0017,0.005$, respectively. Panels in
 the left column show from top to bottom: logarithmic slope of the stellar
 density profile $\gamma_{\star}$; stellar density; anisotropy parameter.
 Panels in the right column show: logarithmic slope of the projected density
 slope profile $\gamma_{\star,\Sigma}$; projected density; line-of-sight 
 velocity dispersion. Black points mark values of $\Rinf$. Densities are 
 normalized to $\rho_{\rh}\equiv\rho(\rh)$, surface densities to 
 $\Sigma_{\Reff}\equiv\Sigma(\Reff)$ and line-of-sight velocity dispersions 
 to $\sigma^{\rm one\ C}_{\rm los, \Reff}$, the line-of-sight
 dispersion of the one component model, computed at $R=\Reff.$
 }\label{fig:STBH}
\end{figure*}

Fig.~\ref{fig:STBH} plots  stellar properties of the models without dark
haloes. The left column plots three three-dimensional diagnostics: from top
to bottom
logarithmic slope $\gamma_{\star}$, density $\rho$ and
anisotropy $\beta$. The right column plots projected quantities: from top to
bottom logarithmic slope $\gamma_{\star,\Sigma}\equiv\text{d}\ln\Sigma/\text{d}\ln R$, 
surface density $\Sigma$ and velocity dispersion $\sigma_{\rm los}$ in 
units of the line-of-sight velocity dispersion at $\Reff$ in the corresponding
one-component model. Solid and dashed lines relate to models with 
$\eta=0.75$ ($\sim$isotropic) and 1 (radially biased), respectively. Values of 
$\muBH$ increase from bottom to top, with orange curves showing models without BHs. 
Black points mark values of $\Rinf$.

It is evident that $\Rinf$ is essentially proportional to $\muBH$ and
insensitive to $\eta$ (Table~\ref{tab:BHtab}). It is also evident that on
the sky the region that is significantly affected by the BH is much smaller
than the corresponding three-dimensional region. In the latter, the stellar
density becomes very cuspy, with $\gamma_{\star}$ approaching $-1.5$ as predicted
by previous works 
\citep{Quinlan1995,BinneyTremaine2008}. 
At $\Rinf$ the logarithmic slope of the 
projected density profile $\gamma_{\star,\Sigma}(\Rinf) \simeq-0.13$ in the
model with the highest $\muBH$. The central divergence of the line-of-sight
velocity dispersion is $\sigma_{\rm los}\simeq r^{-1/2}$, as expected, but sets
in only well inside $\Rinf$. The bottom left panel of Fig.~\ref{fig:STBH}
shows that the models remain isotropic at their centres
\citep{GoodmanBinney}.

\begin{figure*}
 \includegraphics[width=.7\hsize]{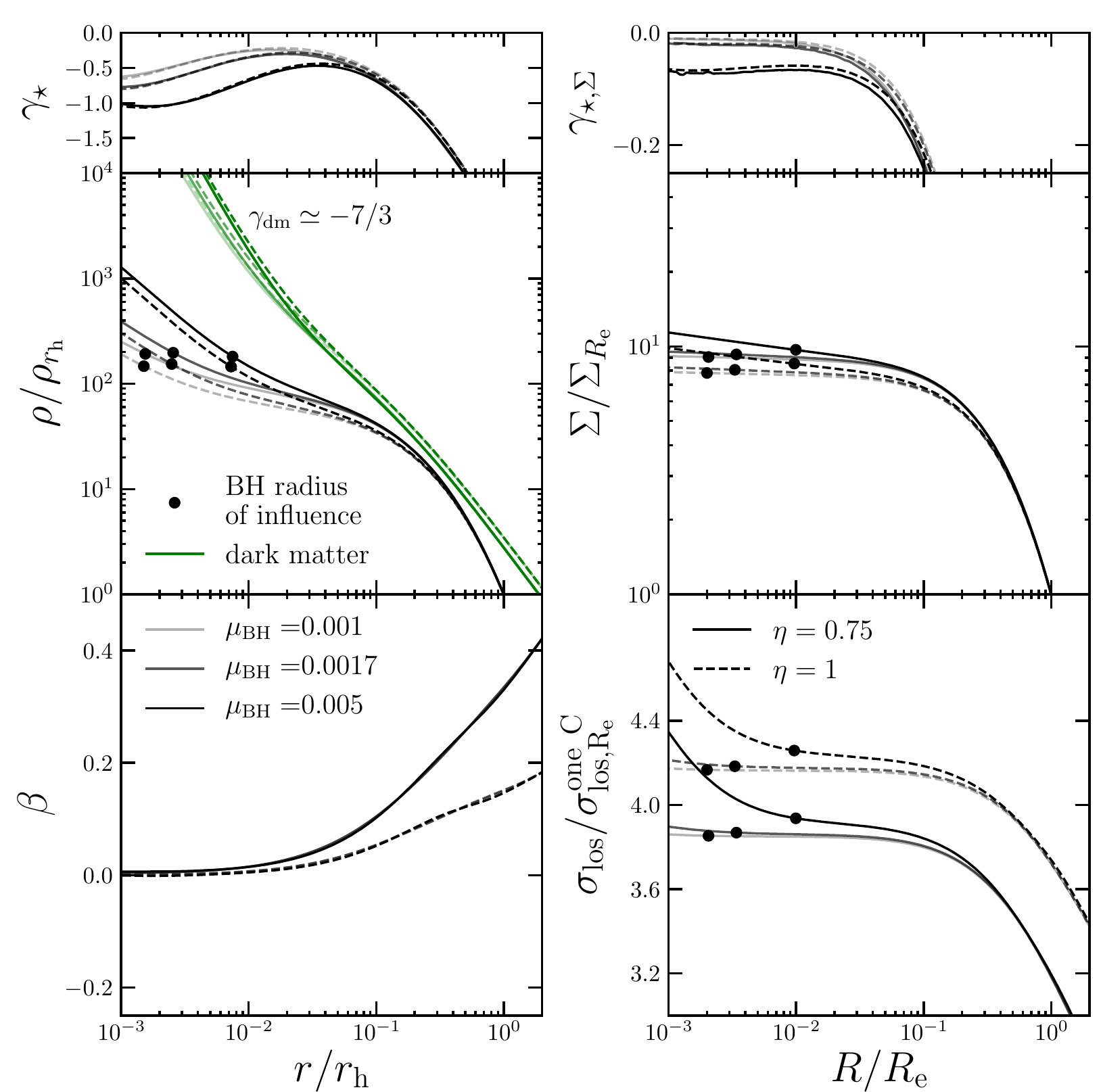}
 \caption{Same as Fig.~\ref{fig:STBH} except for models with an NFW dark
 halo. The green curves in the centre left panel show the density of DM. 
 All grey and black curves refer to the stars.}\label{fig:STDMBH}
\end{figure*}

Fig.~\ref{fig:STDMBH} plots the same quantities as Fig.~\ref{fig:STBH} but
for the models with a dominant dark halo. In a model with both stars and a
dark halo, the slopes of the cusps that the BH creates in each
component are the same ($\rho_{\star}\sim r^{-3/2}$, $\rho_{\rm dm}\sim
r^{-7/3}$, \citealt{Quinlan1995}) 
as those created by BHs in single-component
models. The main effect of adding a dark halo is to increase the stellar
velocity dispersion before addition of a BH, with the consequence
that the dynamical impact of the BH is confined to smaller radii than
in a model without a dark halo; $\Rinf$ shrinks by a factor 2--3
(Table~\ref{tab:BHtab}). The change in the outer stellar velocity
distributions (Fig.~\ref{fig:STDMBH}, bottom panel, left column) 
is only due to different $\Omega_L/\Omega_r$ set by the dark halo 
(see Section~\ref{subsec:NFW}).

\begin{figure*}
 \centering
 \includegraphics[width=.7\hsize]{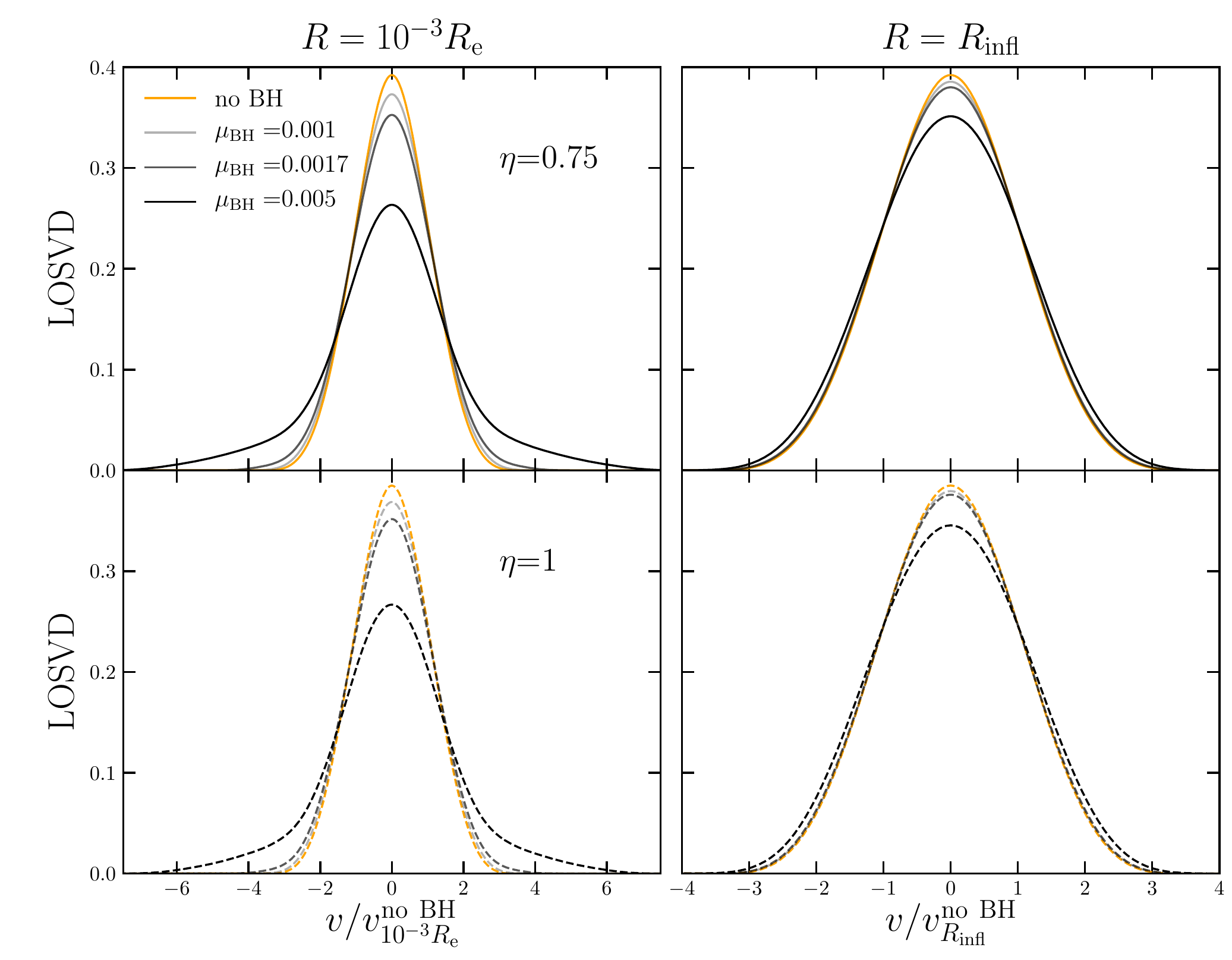}
 \caption{The impact of a BH on LOSVDs in one-component models: upper row
 an isotropic
 model ($\eta=0.75$); lower row a radially biased model ($\eta=1$); left
 column LOSVD at $R=10^{-3}\Reff$; right column LOSVD at $R=\Rinf$. Black
 hole mass fractions from zero to $0.005$ are indicated by line type.
 Velocities are normalized to the line-of-sight velocity dispersion at the
 relevant radius, $v_{R}^{\rm no\, BH}$, with $R=10^{-3}\Reff, \Rinf$, respectively in 
 the left and right columns. The orange curves show the LOSVD of the corresponding model
 with no BH. In this case $\Rinf$ is not defined so we plot the LOSVD at 
 $10^{-2}\Reff$.}\label{fig:losvdBH}
\end{figure*}

Fig.~\ref{fig:losvdBH} shows stellar LOSVDs for models without dark matter,
computed at both $R=\Rinf$ (left column) and $R\simeq10^{-3}\Reff$.
There are substantial differences between the LOSVDs with different $\muBH$ only
at the smaller radius.

These models underline the need for exquisitely accurate surface brightness
profiles and velocity measurements to well inside $\Rinf$ if intermediate massive BHs
(IMBHs) are to be detected in GCs and dSphs. In GCs, $\Rinf$ is often already 
close to the smallest currently resolvable spatial scale. For instance: if $\omega$
Centauri, one of the largest GCs with $\rh\simeq5\,$arcmin
(\citealt{Harris1996}) 
and a good candidate to host an
IMBH \citep{VanDerMarel2010}, 
contained a BH with $\muBH=0.005$, $\Rinf$
would be of the order of 10 arcsec (assuming $\rh\simeq\Reff$, Table~\ref{tab:BHtab}).
Moreover, extreme crowding, the problem of locating the centre of a system, 
and the possibility that any inward increase in the velocity dispersion 
is driven by mass segregation rather than a BH, all make it hard to build a
convincing case for an IMBH in a GC \citep{Zocchi2019}. We have shown that the dark haloes of
dSphs make the problem harder in dSphs by driving $\Rinf$ inwards.

\section{Application to data}
\label{sec:appl}

We have indicated that the DF (\ref{for:df1}) has all the required
features to model the typically observed properties of dSphs and GCs. 
In this Section we justify this statement. 

Fitting models to data for a specific object involves careful consideration
of issues with the data such as degradation by seeing, foreground
contamination, selection effects associated with crowding or field-of-view
limitations and selection of bright stars for spectroscopy. Consequently,
presentation of a thorough fitting exercise of a single system would shift
the focus from the DF (\ref{for:df1}) to the fitted system.  Presentation of
the same exercise for several diverse systems is not feasible in a single
paper. Hence we do not attempt detailed fits. Instead, we plot alongside data
the predictions of a variety of models in the hope of convincing readers that
there are models within the set explored that would provide acceptable fits
to the data after correction of all relevant observational biases.

\subsection{Globular Clusters}
\label{subsec:GC}

\begin{figure*}
 \includegraphics[width=1.\hsize]{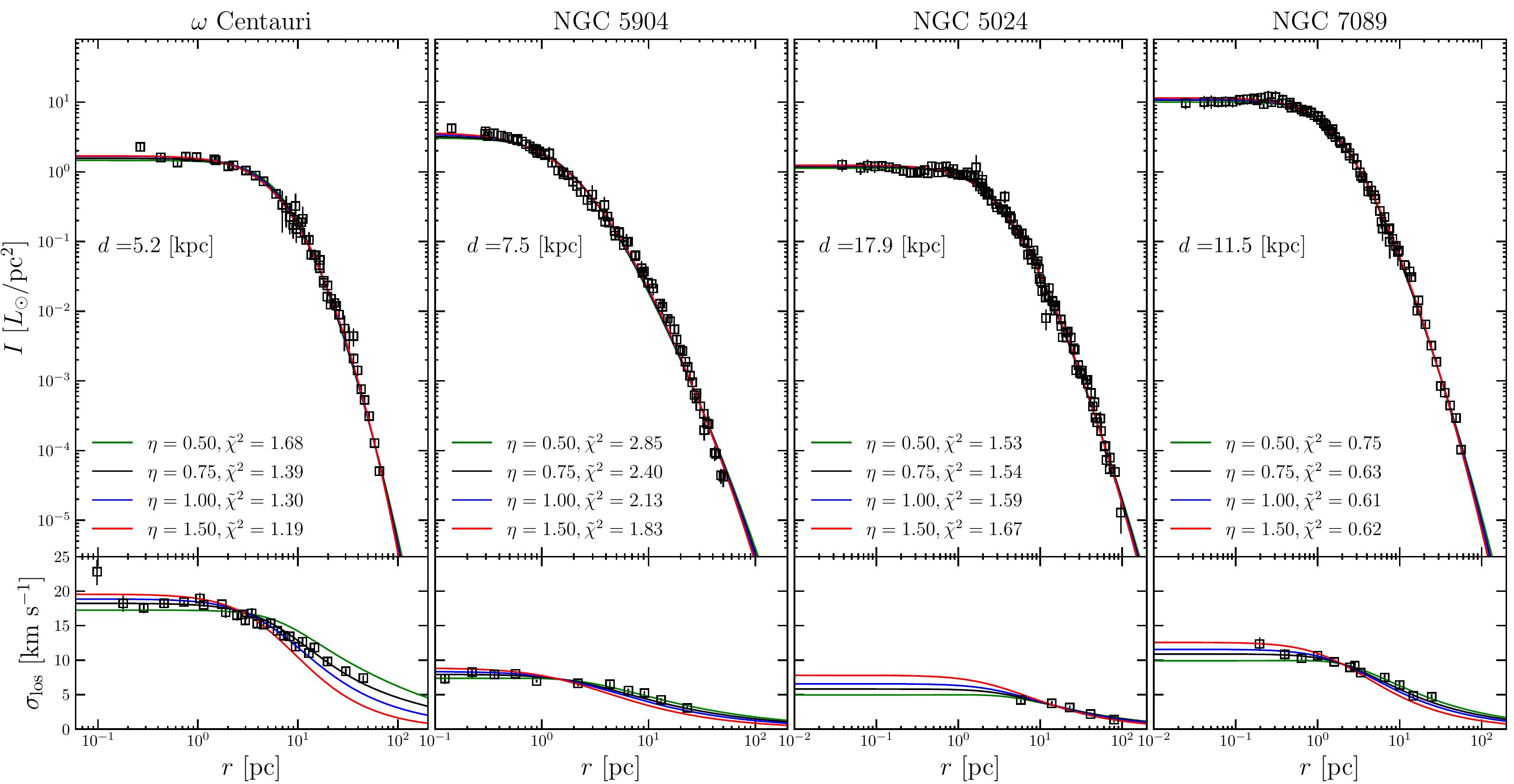}
 \caption{One-component models of globular clusters. From left to right:
 $\omega$ Centauri, NGC 5904, NGC 5024, NGC 7089. The upper panels show 
 data and model predictions for surface brightness. Curves show the 
 models that fit the data best for pre-determined anisotropy: $\eta$=
 0.5 (green), 0.75 (grey), 1 (blue), 1.5 (red). The models' line-of-sight
 velocity dispersion profiles, scaled to match the observed line-of-sight
 velocity dispersion profiles (\citealt{Baumgardt2019}), are shown in the
 bottom panels. For each model we report the value of the reduced chi squared 
 $\tilde{\chi^2}$.}\label{fig:GCs}
\end{figure*}

We chose four representative GCs: $\omega$ Centauri, NGC 5904, NGC 5024 and
NGC 7089. To demonstrate the flexibility of the DF (\ref{for:df1}), 
we fit the surface brightness profiles of each GC with four one-component models, 
each with a different velocity anisotropy. 

Cluster distances are taken from the \cite{Harris1996} catalogue, the surface
brightness profiles from the catalogue of \cite{Trager1995}, while the 
line-of-sight velocity dispersion profiles from \cite{Baumgardt2019}.
The surface brightness data sets consist of triplets of
$\{R_i,I^{\rm obs}_i,\delta I_i\}$, with $i=1,...,N$, where $R_i$ is the 
$i$-th bin's average radius and $I_i$ and $\delta I_i$ are its surface 
brightness and error. The errors are computed following Section 2.2 of
\cite{McLaughlin2005}. The line-of-sight velocity dispersion profiles 
consist of triplets $\{R_k,\sigma_{\rm los, k},\delta \sigma_{\rm los, k}\}$,
with $k=1,...,M$, where $R_k$ is the bin's avarage radius, 
while $\sigma_{\rm los,k}$ and $\delta\sigma_{\rm los, k}$ are its 
line-of-sight velocity dispersion and error, respectively.

We present models with $\eta$ = 0.5, 0.75, 1, 1.5, to cover a wide 
range of anisotropies (see Section~\ref{subsec:oneC}). To determine 
the best fitting model, we minimize the chi squared
\begin{equation}\label{for:chi2}
 \chi^2 \equiv \sum_{n=1}^{N} \biggl(\frac{I^{\rm mod}_i(R_i) 
 - I^{\rm obs}_i}{\delta I_i}\biggr)^2.
\end{equation}
Since equation (\ref{for:chi2}) does not include the fit to the kinematics, 
the only free parameters to be constrained by data are $\alpha$, 
$\rst$ and a normalization parameter $Q\equiv\Mst/\MTL$, where 
$\MTL$ is the mass-to-light ratio. The mass scale $\Mst$ of each model 
is then determined by fitting the observed GC velocity dispersion
profile only. 

Given the few free parameters, we adopt a uniform grid search method to
find the minimum of (\ref{for:chi2}). The model surface brightness 
$I_i^{\rm mod}(R_i)$ is computed assuming a constant mass-to-light 
ratio $\MTL$. The value of $\MTL$ is unambiguously determined by the
requirement that the model provides the total luminosity.\footnote{
Given a model surface brightness profile properly length-scaled, 
the equation 
\begin{equation}
\frac{\partial \chi^2}{\partial Q} = 0
\end{equation}
can be solved analytically.}.

The upper panels of Fig.~\ref{fig:GCs} show that for all four values of
$\eta$ one can fit the very precise photometric data almost perfectly, even
though the data extend over nearly five orders of magnitude in surface
brightness. As measure of the goodness of the fits, Fig.~\ref{fig:GCs} lists 
the values of the reduced chi square, $\tilde{\chi}^2\equiv-\chi^2/$d.o.f., 
where d.o.f. = $N$ - 2. The only slight misfit is at the centre of NGC 5904,
where a mild cusp in the data cannot be reproduced by the DF (\ref{for:df1}). 
The lower panels of Fig.~\ref{fig:GCs} show the line-of-sight velocity
dispersion profiles of the models scaled to match the observed profiles. 
The shape of the line-of-sight velocity dispersion profiles of each GC
is well reproduced by at least one model. The parameters of these models
are listed in Table~\ref{tab:GCs}.

While we have demonstrated that the application of the DF (\ref{for:df1}) to
GCs is promising, our one-component models can only be regarded as starting
points for a much more sophisticated modelling effort. All GCs have
experienced significant mass segregation. Consequently, stars of different
masses and evolutionary stage will be distributed differently in action
space. In particular, more massive stars will be more tightly clustered
towards the origin of action space than less massive stars. Black holes and
neutron stars, will be most tightly clustered around the origin, followed by
horizontal-branch stars, followed by turnoff stars. Low-mass main-sequence
stars will extend furthest from the origin of action space.  Each stellar
type should have its own DF $f(\vJ)$ and be an independent component of a
composite model \citep{GielesZocchi2015,Zocchi2016}. The observables such as surface
brightness and line-of-sight velocity dispersion would be predicted by
weighting these components according to their luminosity. Many GCs show
significant signs of rotation (\citealt{Bianchini2018}), and to reproduce
this aspect of the observations we would need to include in the DF a
component odd in $\Jphi$ (\citealt{Binney2014}, see also
\citealt{Jeffreson2017} who used a different family of action-based DFs
to reproduce flattened, rotating and almost isotropic GCs).

\begin{table}
 \begin{center}
  \caption{Parameters of the models fitted to GC data. $\eta$ and 
  $\alpha$ are dimensionless parameters in the DF (\ref{for:df1}). 
  $\Jst$ is the action scale, while $\Mst$ is the total mass.
  $\Mst/\LV$ is the mass-to-light ratio, with $\LV$ the total
  luminosity in the V band, taken from \citet{Harris1996}. $\chi^2$
  of the best-fitting model is defined by equation (\ref{for:chi2}).
  $N$ is number of bins in the observed surface brightness profile.}\label{tab:GCs}  
  \begin{tabular}{lccccc}
  \hline\hline
  \multicolumn{6}{c}{$\omega$ Centauri ($N$=51)} \\  
  \hline\hline  
  $\eta$&$\alpha$&$\Jst$ [kpc km s$^{-1}$]	& $\Mst$ [$10^5 M_{\odot}$] & $\Mst/\LV$ & $\chi^2$ \\
  0.5	&$0.931$ &2.15	& 34.8 & 3.20   &84.51  \\
  0.75  &$0.954$ &2.87  & 31.7 & 2.91   &68.22  \\
  1     &$1.02$  &3.78  & 29.2 & 2.69   &63.88   \\
  1.5   &$1.26$  &6.59  & 26.3 & 2.42   &58.40   \\
  \hline\hline
  \multicolumn{6}{c}{NGC 5904 ($N$=78)} \\
  \hline\hline
  $\eta$&$\alpha$&$\Jst$ [kpc km s$^{-1}$]	& $\Mst$ [$10^5 M_{\odot}$] & $\Mst/\LV$ & $\chi^2$ \\
  0.5 &0.503 &$6.70\times10^{-2}$ &  3.06 & 1.07 & 216.92  \\
  0.75&0.522 &$1.08\times10^{-1}$ &  2.98 & 1.04 & 182.09  \\
  1   &0.543 &$1.55\times10^{-1}$ &  2.88 & 1.00 & 161.60  \\
  1.5 &0.605 &$3.15\times10^{-1}$ &  2.72 & 0.95 & 139.98  \\
  \hline\hline
  \multicolumn{6}{c}{NGC 5024 ($N$=111)} \\   
  \hline\hline
  $\eta$&$\alpha$&$\Jst$ [kpc km s$^{-1}$]	& $\Mst$ [$10^5 M_{\odot}$] & $\Mst/\LV$ & $\chi^2$ \\
  0.5 &0.464& $1.25\times10^{-1}$  & 2.54  & 0.98 & 166.82   \\
  0.75&0.480& $2.18\times10^{-1}$  & 2.94  & 1.13 & 168.15   \\
  1   &0.502& $3.60\times10^{-1}$  & 3.30  & 1.27 & 172.87   \\
  1.5 &0.556& $8.50\times10^{-1}$  & 3.96  & 1.52 & 181.53   \\
  \hline\hline
  \multicolumn{6}{c}{NGC 7089 ($N$=82)} \\
  \hline\hline
  $\eta$&$\alpha$&$\Jst$ [kpc km s$^{-1}$]	& $\Mst$ [$10^5 M_{\odot}$] & $\Mst/\LV$ & $\chi^2$ \\
  0.5  &0.500 &$2.29\times10^{-1}$&  7.83 & 2.24 & 60.34 \\
  0.75 &0.517 &$3.62\times10^{-1}$&  7.85 & 2.24 & 50.45 \\
  1    &0.540 &$5.55\times10^{-1}$&  7.90 & 2.26 & 48.77 \\
  1.5  &0.600 &1.14               &  7.84 & 2.24 & 49.58 \\
  \hline\hline
  \end{tabular}
 \end{center}
\end{table}

\subsection{Dwarf spheroidal galaxies}
\label{subsec:dSphs}

\cite{Pascale2018} demonstrated that the DF (\ref{for:df1}) yields very
accurate models of the Fornax dSph. Here we model five further dSphs: 
Carina, Leo I, Sculptor, Sextans and Ursa Minor, with the aim to prove
that the use of the DF (\ref{for:df1}) can be extended to the whole
population of classical dSphs. We present spherical, anisotropic
models, with separate DFs for the stellar and the halo components, which 
just fit the dSph number density profiles, given a certain orbital anisotropy.
For Sculptor we present three-component models, which have distinct DFs 
for the red and blue horizontal branch stars and the dark matter halo.


The projected number density profiles of the Carina, Leo I, Sextans and 
Ursa Minor dSphs have been taken from \cite{Irwin1995}, while their 
line-of-sight velocity dispersion profiles are from \cite{Walker2007}. 
The projected number density and line-of-sight velocity dispersion profiles
of the distinct populations of Sculptor are from \cite{Battaglia2008}. 
We adopt distances from \cite{Mateo1998}.

\subsubsection{Carina, Leo I, Sextans and Ursa Minor}

Our analysis proceeds essentially as described in Section~\ref{subsec:GC}.
The photometric contribution is now computed from triplets $\{R_i, 
n_{\star, i}^{\rm obs}, \delta n_{\star, i}^{\rm obs}\}$, where 
$ n_{\star, i}^{\rm obs}$ and $\delta n_{\star, i}^{\rm obs}$ are a
number density and its error. The predicted number density, $n_{\star}^{\rm mod}$,
is computed from the surface density of mass assuming a constant mass per
detected star, $\overline{m}$. The kinematics is computed from triplets  
$\{R_k,\sigma_{{\rm los},k},\delta\sigma_{{\rm los},k}\}$.
The stellar component of each dSph is represented by DF (\ref{for:df1}),
with fixed stellar masses $\Mst$ (see Table~\ref{tab:dSphs}). The dark matter halos are described 
by the cuspy DF (\ref{for:dmdf}, $\Jctil=0$). For each dSph with stellar mass $\Mst$,
according to estimetes of the low mass end of the stellar-to-halo mass 
relation \citep{Read2017}, and to the halo-mass concentration ralation
\citep{Munoz2011}, we fix the dark matter mass enclosed within the halo 
scale radius $\Mdm(<\rsdm)$, and the halo scale radius $\rsdm$, to values 
predicted by cosmology. The prescribed values of $\Mdm(<\rsdm)$ and $\rsdm$
are obtained by varying iteratively $\Mdm$ and $\Jdm$ (the final values 
of these parameters are given in Table~\ref{tab:dSphs}). 

\begin{figure*}
 \includegraphics[width=1\hsize]{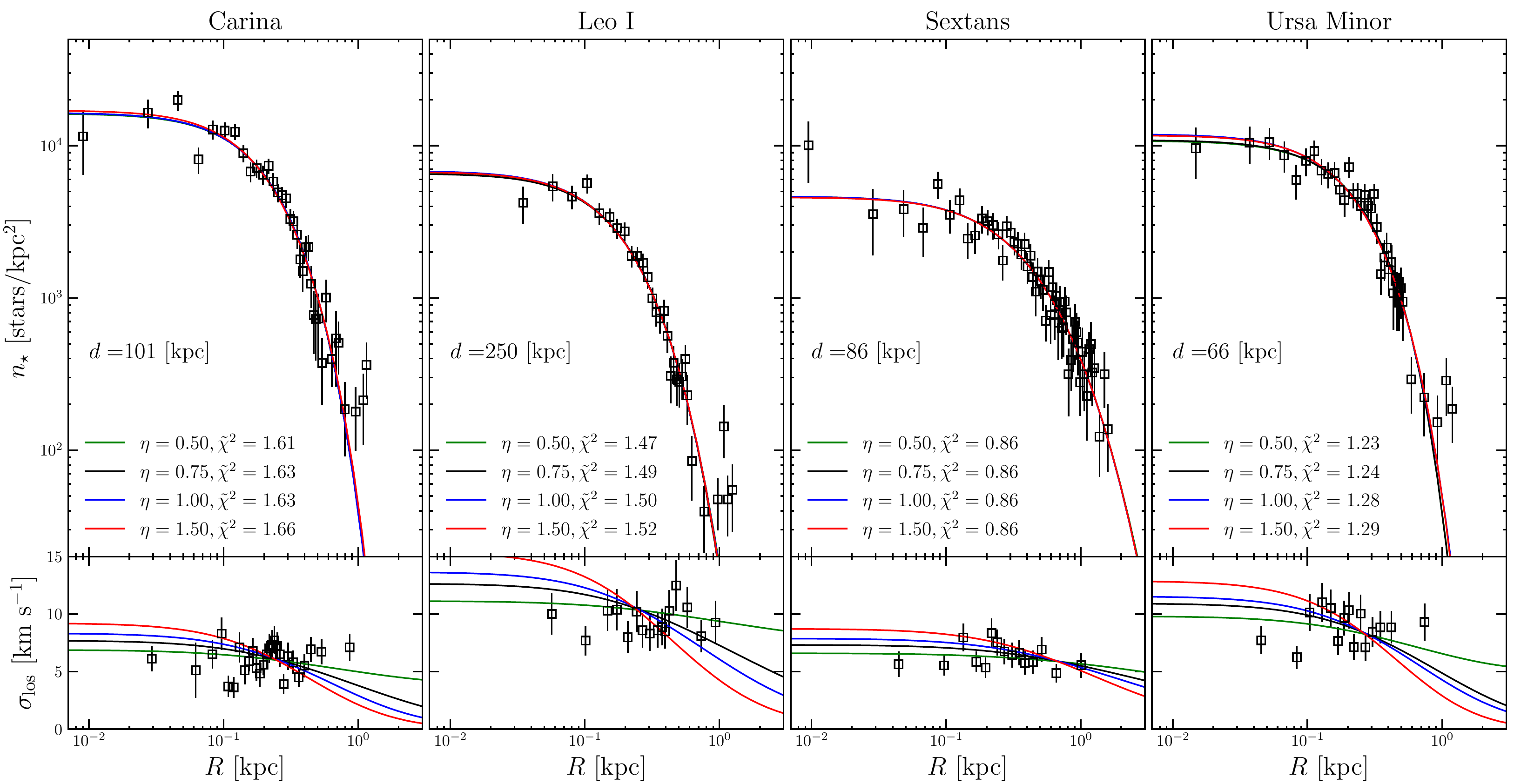}
 \caption{Models of dSphs with a stellar comonent and a dark matter halo. 
 Columns from the left to right: Carina, Leo I, Sextans, Ursa Minor. Upper 
 panels show projected number densities together with fits by four models with
 pre-determined stellar velocity anisotropy: $\eta$= 0.5 (green), 0.75 (grey), 1 (blue),
 1.5 (red). The parameters of the models are reported in Table~\ref{tab:dSphs}. 
 Lower panels show observed line-of-sight velocity dispersions
 and  the models' predictions. For each model we report the value of the
 reduced chi squared.}\label{fig:dSphs}
\end{figure*}

The upper panels of Fig.~\ref{fig:dSphs} show that for all four values of
$\eta$ the best  DF provides an excellent fit to the observed number density
profiles of the four galaxies. The lower panels shows the observed velocity
dispersion profiles of the galaxies alongside the predictions for each value
of $\eta$.

Table~\ref{tab:dSphs} gives the values of the parameters and of $\chi^2$ for
the best-fitting models of Carina, Leo I, Sextans and Ursa Minor. It also
gives the parameters and $\chi^2$ for the best-fitting \cite{Sersic1968}
profile
\begin{equation}\label{for:Sersic}
 n_S(R) = n_0 \exp\biggl[-\biggl(\frac{R}{R_S}\biggr)^{1/m}\biggr].
\end{equation}
Every DF yields a comparable or lower $\chi^2$ than does the S\'ersic profile. 
This is remarkable in as much as (i) fits of both the DF and the 
S\'ersic profile require searches over just two parameters in 
addition to a basic scaling parameter, yet (ii) the DF defines 
a complete, dynamically consistent six-dimensional model whereas
the S\'ersic profile provides nothing beyond the radial run of density. 
Consequently, it can be argued that a dSph is more effectively described 
by the parameters of its best-fitting $\fJ$ than by the parameters of
the best-fitting S\'ersic profile.

\subsubsection{Sculptor}

\begin{figure*}
\centering
 \includegraphics[width=.8\hsize]{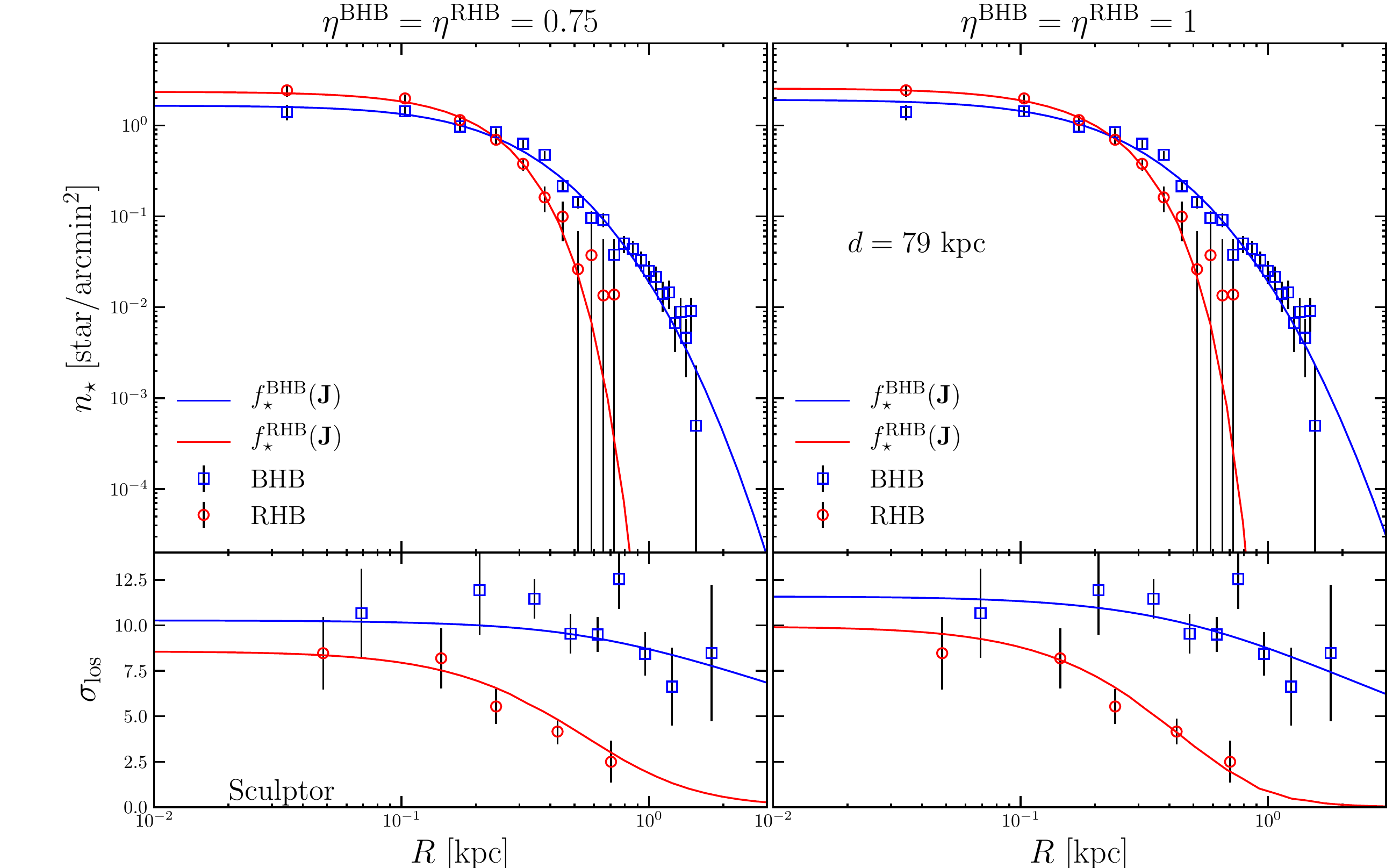}
 \caption{ Three-component models of the Sculptor dSph (two stellar components and a dark matter halo).
 The velocity anisotropy of the stellar components is slightly radially biased
 ($\eta=0.75$) in the left column and radially biased ($\eta=1$) in the right
 column. Red and blue curves in the upper panels show model fits to the observed
 surface densities of RHB and BHB, respectively. The lower panels show the
 predicted line-of-sight velocity dispersion profiles of each component alongside 
 the observed profiles.}\label{fig:Sculp}
\end{figure*}

dSphs usually exhibit complex star formation and chemical enrichment
histories. These galaxies seem to experience bursts of star formation, and
the stars formed in each burst are distributed differently in action space.
Since all populations move in a common potential, observations that are
able to distinguish between the populations have the potential to
constrain the system's gravitational field more strongly than is possible in
a system with only a single population
\citep{WalkerPen2011,Agnello2012,Amorisco2013}.

We model two populations in the Sculptor dSph, with each population  
described by the DF (\ref{for:df1}), and with a separate component 
describing a dark matter halo DF (\ref{for:dmdf}). The two populations
are the stars on the blue (red) horizontal branch BHB (RHB), which are less
(more) metal rich and more (less) extended spatially.

We will refer to all the parameters belonging to the BHB (RHB)  
populations, as $\ast^{\rm BHB}$ ($\ast^{\rm RHB}$) where $\ast$ = 
$\alpha$, $\eta$, $\Mst$, $\Jst$. For simplicity in each model $\eta$  
is the same for both populations and the total stellar mass
$\MstBHB+\MstRHB$ is fixed. We consider two representative cases,
$\eta=0.75$ (slightly radially biased) and $\eta=1$ (radially biased). 
We assume a cored dark matter halo described by DF (\ref{for:dmdf};
$\Jctil=0.02$). As for the other dSphs, we fix the enclosed mass $\Mdm(<\rsdm)$
and the scale radius $\rsdm$ to cosmologically motivated values.

Then the model's free parameters are
\begin{equation}\label{for:scparam}
\xib = \biggl(\alpha^i,\Jst^i,\frac{\MstBHB}{\MstBHB+\MstRGB},\Jdm,\Mdm\biggr),
\quad (i=\hbox{RHB, BHB}).
\end{equation}

We minimize the figure of merit
\begin{equation}
 \chi^2_{\rm tot} = \chiRHB + \chiBHB,
\end{equation}
where $\chi^2$ for each population is defined by equation (\ref{for:chi2}).

In view of the higher dimensionality of this problem, we explored the 
parameter space using a stochastic search method based on a Markov-Chain
Monte Carlo (MCMC) algorithm, with a Metropolis-Hastings (\citealt{Metropolis1953},
\citealt{Hastings1970}) sampler, to sample from the posterior distribution.
We used uninformative, flat priors on the free parameters (\ref{for:scparam}).

In the upper panels of Fig.~\ref{fig:Sculp} squares and circles mark the 
number densities of BHB and RHB stars, respectively. The predictions for
these populations of the best-fitting models are shown by blue and red curves, 
respectively. The left panel shows the fit provided by the mildly radially 
biased model, and the right panel shows the fit provided by the radially biased
model. It is clear that both three-component models provide excellent fits to
the data, and that also the models predictions on the line-of-sight velocity 
dispersion profiles provide an excellent description of the data. 
Table \ref{tab:Sculp} gives the models' parameters. 

These simple test cases prove that the extension of the DF (\ref{for:df1}) to the 
whole system of classical dSphs is possible and promising, whether the galaxy 
is represented as a single stellar population or in more sophisticated model
that reflects the chemodynamic history of the system.

\begin{table*}
 \begin{center}
  \caption{Parameters of two-component models fitted to dwarf spheroidal 
  galaxies.$\eta$ and $\alpha$ are dimensionless parameters in the DF 
  (\ref{for:df1}). $\Jst$ is the action scale while $\Mst$ is the dSph 
  total mass: $^1$ \citet{Ural2015}, $^2$ \citet{Weisz2014}, $^3$ 
  \citet{Karlsson2012}. $\Mdm$ and $\Jdm$ are the halo total mass and 
  action scale equation (\ref{for:dmdf}). The dark halo DF is cuspy, with 
  $\Jctil=0$. The figure of merit $\chi^2$ of the best-fitting model is
  defined by equation (\ref{for:chi2}). $N$ is number of bins in the
  observed star-count profile. $n_0$, $m$ and $R_S$ are the normalization, 
  S\'ersic index and scale radius, respectively, of the best-fitting 
  S\'ersic profile (equation~\ref{for:Sersic}).}\label{tab:dSphs}
  \begin{tabular}{ccccccccccc}  
  \hline\hline
  \multicolumn{10}{c}{Carina ($N$=36)} 	\\
  \multicolumn{6}{c}{$\fJ$ model} &	\multicolumn{4}{c}{S\'ersic Fit}	\\
  \hline\hline  
  $\eta$&$\alpha$&$\Jst$ [kpc $\kms$]	&$\Mst$	[10$^6 \Msun$]& $\Mdm$ [10$^8 \Msun$]&  $\Jdm$ [kpc $\kms$] & $\chi^2$  &  $n_0$ $[n_{\star}$ kpc$^{-2}]$	&   $m$    & $R_s$ [kpc]	& $\chi^2$	\\
  0.5	&0.946 &0.677 & 0.48$^1$  & 8.69  &  44.58    & 56.42      & 14.17	&0.813  & 0.215	    & 57.42	\\ 
  0.75  &1.10  &1.21  &           &       &           & 56.88      \\
  1     &1.33  &1.96  &           &       &           & 57.11      \\
  1.5   &1.81  &3.48  &           &       &           & 57.90      \\
  \hline\hline
  \multicolumn{10}{c}{Leo I ($N$=31)} \\
  $\eta$&$\alpha$&$\Jst$ [kpc $\kms$]	&$\Mst$	[10$^6 \Msun$]& $\Mdm$ [10$^9 \Msun$]&  $\Jdm$ [kpc $\kms$] & $\chi^2$  &  $n_0$ $[n_{\star}$ kpc$^{-2}]$	&   $m$    & $R_s$ [kpc]	& $\chi^2$	\\
  \hline\hline
  0.5 	&0.714 & 0.513 & 5.5$^2$  &  6.57  &  174.3  & 44.03      	& 38.03 &0.876  &  0.182    & 44.80 \\
  0.75	&0.860 & 1.18  &          &        &         & 44.63      \\
  1   	&0.933 & 1.74  &          &        &         & 44.87      \\
  1.5 	&1.34  & 4.20  &          &        &         & 45.52      \\
  \hline\hline
  \multicolumn{10}{c}{Sextans ($N$=56)} \\
  $\eta$&$\alpha$&$\Jst$ [kpc $\kms$]	&$\Mst$	[10$^6 \Msun$]& $\Mdm$ [10$^8 \Msun$]&  $\Jdm$ [kpc $\kms$] & $\chi^2$  &  $n_0$ $[n_{\star}$ kpc$^{-2}]$	&   $m$    & $R_s$ [kpc]	& $\chi^2$	\\
  \hline\hline
  0.5 	&0.594	& 0.420 & 0.5$^3$  &  7.94  &  47.0  & 47.31    	& 3.33  & 1.13  &  0.339   & 48.54 \\
  0.75	&0.656	& 0.828 &          &        &        & 47.39 	\\ 
  1   	&0.724	& 1.41  &          &        &        & 47.40 	\\
  1.5 	&0.902	& 3.33  &          &        &        & 47.33 	\\
  \hline\hline
  \multicolumn{10}{c}{Ursa Minor ($N$=37)} \\
  $\eta$&$\alpha$&$\Jst$ [kpc $\kms$]	&$\Mst$	[10$^6 \Msun$]& $\Mdm$ [10$^9 \Msun$]&  $\Jdm$ [kpc $\kms$] & $\chi^2$  &  $n_0$ $[n_{\star}$ kpc$^{-2}]$	&   $m$    & $R_s$ [kpc]	& $\chi^2$	\\
  \hline\hline
  0.5  &1.17 & 1.54 & 0.29$^2$ &  1.316  &  50.7  &  44.43		& 3.62 & 0.665 & 0.278 & 42.743 \\
  0.75 &1.39 & 2.61 &          &         &        &  44.82 	\\
  1    &1.32 & 2.94 &          &         &        &  46.00 	\\
  1.5  &2.20 & 6.03 &          &         &        &  46.42  	\\
  \hline\hline
  \end{tabular}
 \end{center}
\end{table*}

\begin{table*}
 \begin{center}
  \caption{Parameters of DFs fitted to three components the Sculptor dwarf
  spheroidal galaxy. $\eta^{\rm pop}$ and $\alpha^{\rm pop}$ are dimensionless
  parameters in the DF (\ref{for:df1}). $\Jst^{\rm pop}$ is the scale action defined
  by the DF. $\Mst^{\rm pop}$ the component's mass. $\Jdm$ and $\Mdm$ are 
  the halo action scale and total mass (equation \ref{for:dmdf}). The dark halo DF 
  (\ref{for:dmdf}) is cored, with $\Jctil=0.02$. The figure of 
  merit $\chi^2$ of the best-fitting model is defined by equation 
  (\ref{for:chi2}). The BHB and RHB star-count profiles have 
  a number of bin $N^{\rm BHB}=23$ and $N^{\rm RHB}=11$, respectively.
  The total stellar mass $\MstBHB+\MstRHB = 2.3\times10^6\Msun$ \citealt{Weisz2014}.}\label{tab:Sculp}
  \begin{tabular}{lcccccccc}  
  \hline\hline
  \multicolumn{9}{c}{Sculptor}\\
  \hline\hline  
  $\etaBHB = \etaRHB$ & $\alphaBHB$ & $\JstBHB$ [kpc $\kms$] & $\alphaRHB$ & $\JstRHB$ [kpc $\kms$] & $\frac{\MstBHB}{\MstBHB+\MstRHB}$ & $\Jdm$ [kpc $\kms$] & $\Mdm$ [$10^9\Msun$] & $\chi^2$\\
  \hline\hline
  0.75	&  0.591  &  0.389  &  1.83  &  1.86  &  0.892  &  148.2  &  5.87  & 49.50  \\
  1	&  0.554  &  0.359  &   2.41 &  2.79  &  0.736  &  167.8  &  7.36  & 46.09  \\   
  \hline\hline
  \end{tabular}
 \end{center}
\end{table*}

\section{Conclusions}
\label{sec:conc}

As we acquire more complete data for galaxies and star clusters, more
sophisticated models are required to fit the data well and to provide
predictions for further observations that can be tested by extending the
available data. Full exploitation of the best current data requires models
that (i) include several components and (ii) predict not just velocity
moments but full LOSVDs. Models that meet these criteria are readily
constructed if we use action integrals as the arguments of the DF. A
self-consistent model that provides a good fit to a given system can be
quickly constructed by allocating each component, disc, stellar halo, dark
halo, etc., a DF with an appropriate functional form. In this paper we have
explored the scope of the DF (\ref{for:df1}) that was introduced by
\cite{Pascale2018} to model the Fornax dSph. This  DF complements DFs
previously introduced by 
\cite{Binney2010} and \cite{Posti2015} in yielding spheroidal systems
with exponential density profiles.  

The DF has two key parameters, $\eta$ and $\alpha$, which principally control
velocity anisotropy and the radial density profile, respectively. We have
explored models that contain only stars and models that also have a dark halo. We
have investigated the impact that the dark halo has on stellar observables
both when the halo has been adiabatically distorted by the stars from the
classic NFW form, and when dark-matter particles have been scattered out of
low-action orbits to form a dark core. We have also explored
models in which a massive BH sits at the centre of the galaxy.

We have shown that models generated by the \cite{Pascale2018} DF provide
excellent fits to both globular clusters and to four dSph galaxies. 
The surface-brightness profiles can be fitted equally well with
models that have a wide range of velocity anisotropies, from radially to
tangentially biased.  These models provide an extremely convenient platform
from which to explore that potential of observations to detect dark matter
and IMBHs in globular clusters or dSphs. We have also presented a
three-component model of the Sculptor dSph that describes perfectly the
different spatial extents of the stars on the blue and red horizontal
branches, again for a wide range of assumed velocity anisotropies.

The models presented are all been non-rotating and spherical. One of
the strengths of the $f(\vJ)$ modelling technique is the ease with which a
spherical model can be flattened and set rotating \citep{Binney2014}, and a
forthcoming paper will explore rotating and flattened models systematically

\section*{Acknowledgments}

RP thanks the Rudolf Peierls Centre for Theoretical Physics 
in Oxford for the hospitality 
during the period in which this work has been carried out.
JB acknowledges support from the UK Science and Technology
Facilities Council under grant number ST/N000919/1.
LP acknowledges financial support from a VICI grant from the 
Netherlands Organization for Scientific Research (NWO).
We thank P. Das, J. Magorrian, R. Sch\"onrich and E. Vasiliev 
for very helpful discussions and suggestions, and G. Battaglia 
for sharing observational data.


\bibliographystyle{mnras}
\bibliography{Pascale_expDF}

\end{document}